\documentclass[preprint,  review, 3p]{elsarticle}
\usepackage[x11names]{xcolor}
\usepackage[colorlinks=true]{hyperref}
\AtBeginDocument{\hypersetup{allcolors=DodgerBlue4}}
\usepackage{amssymb}
\usepackage{amsmath,bm}
\PassOptionsToPackage{unicode}{hyperref}
\PassOptionsToPackage{naturalnames}{hyperref}
\usepackage{xfrac}
\usepackage{enumitem}

\usepackage{multirow}
\usepackage{tikz}
\usetikzlibrary{shapes,arrows}
\tikzstyle{block} = [rectangle, draw,
    text width=5em, text centered, rounded corners, minimum height=4em]
\tikzstyle{line} = [draw, -latex']    
\tikzstyle{optional}=[dashed,fill=gray!50]
\tikzstyle{VineNode} = [ellipse, fill = white, draw = black, text = black, align = center, minimum height = 1cm, minimum width = 1cm]
\tikzstyle{DummyNode}  = [draw = none, fill = none, text = black] 
\tikzstyle{TreeLabels} = [draw = none, fill = none, text = black] 
\newcommand{\xshiftNodes}{0.7*\linewidth}
\newcommand{\yshiftLabels}{-.25cm}  
\newcommand{\labelsize}{\footnotesize} 
\newcommand{\xshiftTree}{0.5cm}    

\usetikzlibrary{arrows.meta,chains,positioning,shapes.geometric}

\usepackage{amsthm}
\newtheorem{example}{Example}[section]
\newcommand{\R}{\mathbb{R}}
\usepackage{subcaption}
\DeclareMathOperator*{\argmax}{arg\,max}

\usepackage{algorithm}
\usepackage[noend]{algpseudocode}
\makeatletter
\def\BState{\State\hskip-\ALG@thistlm}

\newcommand*{\rom}[1]{\expandafter\@slowromancap\romannumeral #1@}
\usepackage{lineno}
\makeatother

\algdef{SE}[SUBALG]{Indent}{EndIndent}{}{\algorithmicend\ }%
\algtext*{Indent}
\algtext*{EndIndent}
\usepackage{float}
\usepackage{comment}
\usepackage{xcolor}

\journal{Elsevier}

\biboptions{authoryear}

\begin{document}

\begin{frontmatter}

\title{Vine copula mixture models and clustering for non-Gaussian data}


\author[mysecondaryaddress]{Özge Sahin\corref{mycorrespondingauthor}}
\ead{ozge.sahin@tum.de}
\author[mysecondaryaddress,myaddress]{Claudia Czado}
\cortext[mycorrespondingauthor]{Corresponding author}
\ead{cczado@ma.tum.de}

\address[mysecondaryaddress]{Department of Mathematics, Technische Universität München, Boltzmanstraße 3, 85748 Garching, Germany}
\address[myaddress]{Munich Data Science Institute, Walther-von-Dyck-Straße 10, 85748 Garching, Germany}

\begin{abstract}
The majority of finite mixture models suffer from not allowing asymmetric tail dependencies within components and not capturing non-elliptical clusters in clustering applications. Since vine copulas are very flexible in capturing these dependencies, a novel vine copula mixture model for continuous data is proposed. The model selection and parameter estimation problems are discussed, and further, a new model-based clustering algorithm is formulated. The use of vine copulas in clustering allows for a range of shapes and dependency structures for the clusters. The simulation experiments illustrate a significant gain in clustering accuracy when notably asymmetric tail dependencies or/and non-Gaussian margins within the components exist. The analysis of real data sets accompanies the proposed method. The model-based clustering algorithm with vine copula mixture models outperforms others, especially for the non-Gaussian multivariate data.
\end{abstract}


\end{frontmatter}

\section{Introduction}\label{intro}
Finite mixture models are convenient statistical tools for model-based clustering. They assume that observations in the multivariate data can be clustered using $k$ components. Each component has its density, and each observation is assigned to a component with a probability. They have many applications in finance, genetics, and marketing (e.g., \cite{Hu2006}; \cite{Gambacciani2017}; \cite{Sun2017}; \cite{Zhang2017}). \cite{McLachlan2000} provides more details about the finite mixture models. Both \cite{Bouveyron2014} and \cite{Mcnicholas2016} review recent model-based clustering methods.

One of the main questions to be addressed in the finite mixture models is \textit{how to select the density of each component}. An early answer to this question is to assume a symmetric distribution such as multivariate normal distribution (e.g., \cite{Celeux1995}; \cite{Fraley1998}) or multivariate t distribution (e.g., \cite{Peel2000}; \cite{Andrews2011}). However, these models cannot accommodate the shape of asymmetric components. \cite{Hennig2010} showed one such data example. Their skewed formulations and factor analyzers, therefore, have been extensively studied (e.g., \cite{Lin2007}; \cite{Lee2014}; \cite{Murray2017}).

Additionally, the models based on other distributions, for example, shifted asymmetric Laplace distributions (\cite{Franczak2014}), multivariate power exponential distributions (\cite{Dang2015}), and generalized hyperbolic distributions (\cite{Browne2015}) have been proposed for the past few years. Since one of the main interests of copulas is to relax the normality assumption both in marginal distributions and dependence structure, the finite mixture models with copulas have also been studied (e.g., \cite{Diday2005}; \cite{Kosmidis2016}; \cite{Zhuang2021}). \cite{Cuvelier2002} worked with the Clayton copula to represent lower tail dependence within the components, while \cite{Vrac2005} applied the Frank copula to have non-Gaussian but symmetric dependence. Nevertheless, these methods raise another question in the finite mixture models: \textit{how to select flexible densities of each component} so the model can represent different asymmetric or/and tail dependencies for different pairs of variables. For this question, the vine copula or pair copula construction is a flexible and efficient tool in high-dimensional dependence modeling (\cite{Joe1996}; \cite{Aas2009}).

\begin{figure}[ht]
        \centering
        \begin{subfigure}[b]{0.3\textwidth}
            \centering
            \includegraphics[width=\textwidth]{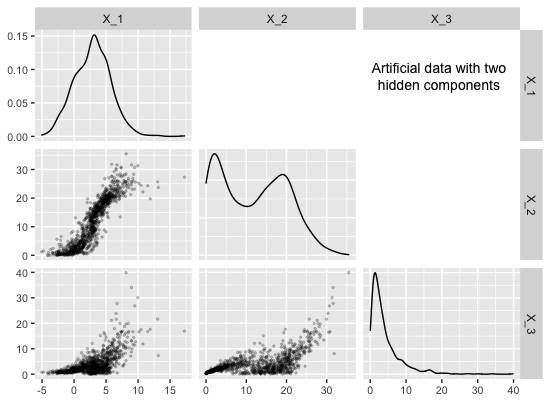}
        \end{subfigure}
        \hspace{2cm}
        \begin{subfigure}[b]{0.3\textwidth}  
            \centering 
            \includegraphics[width=\textwidth]{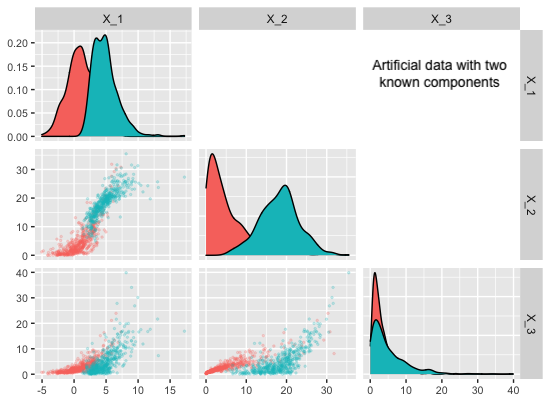}
        \end{subfigure}
        \vskip\baselineskip
        \begin{subfigure}[b]{0.3\textwidth}   
            \centering 
            \includegraphics[width=\textwidth]{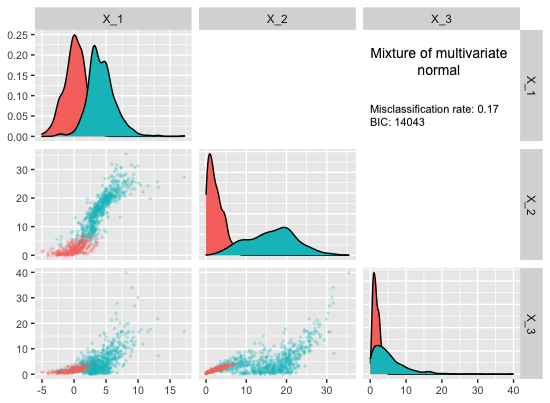}
        \end{subfigure}
        \hspace{2cm}
        \begin{subfigure}[b]{0.3\textwidth}   
            \centering 
            \includegraphics[width=\textwidth]{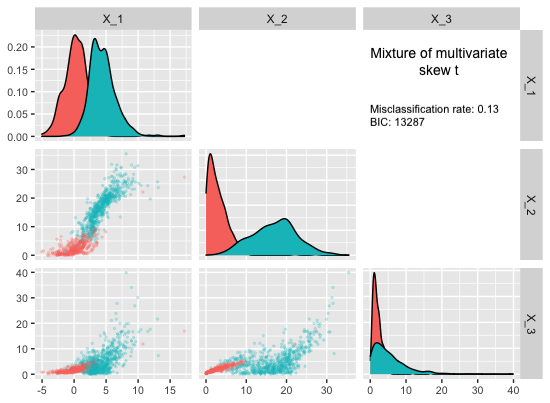}
        \end{subfigure}
   \caption{Pairwise scatter plot of a simulated data set ($500$ observations per component) under the scenario specified in Figure \ref{fig:vcmm-case-1} and Table \ref{table:vcmm-case-1} (top left). The red and green points show the observations of one cluster and the other cluster (top right), respectively. The bottom left and bottom right plots display the fitted mixture of multivariate normal and multivariate skew t distributions, respectively. The diagonal of the plots gives the fitted marginal density function for each component.}
   \label{fig:intro}
\end{figure}

For illustration, consider a data set simulated from a mixture of two three-dimensional vine copulas shown in Figure \ref{fig:intro}. Its data generating process includes asymmetric tail and non-Gaussian dependencies in the components. Most univariate margins are chosen to be non-Gaussian and heavy-tailed. As seen in the top left panel of Figure \ref{fig:intro}, the components are well separated on two out of three bivariate scatter plots. Marginal multimodality can be seen for the second variable. The resulting components are non-elliptical, similar to a banana shape, as shown in the top right. After fitting a mixture of multivariate normal distributions with two components, its challenge to capture the true shape of the components can be seen in the bottom left panel, where the associated Bayesian Information Criterion (BIC) \citep{Schwarz1978} and misclassification rate are provided. Even though fitting a mixture of multivariate skew t distributions with two components in the bottom right provides a better fit than the mixture of multivariate normal distributions fit, it can also not reveal the true characteristics of the data set. The mixture of multivariate normal distributions and multivariate skew t distributions is fitted using the {\fontfamily{pcr}\selectfont R} packages {\fontfamily{pcr}\selectfont mclust} \citep{Scrucca2016} and {\fontfamily{pcr}\selectfont mixsmsn} \citep{mixsmsn}, respectively. The fitting procedures apply 100 different seeds and report the best performance in terms of the misclassification rate.

To capture such behavior, applying the vine copula in finite mixture models has been explored before (e.g., \cite{Kim2013}; \cite{Roy2014}; \cite{Weiss2015}; \cite{Sun2017}). However, they only worked with a subclass of vine tree structures and a small number of pair copula families. Therefore, a vine copula mixture model with all classes of vine tree structures and many different pair copula families is needed. Since it provides flexible densities, formulating its model-based clustering algorithm overcomes the drawbacks mentioned above, especially for non-Gaussian data.

In this paper, we formulate a vine copula mixture model for continuous data allowing all types of vine tree structures, parametric pair copulas, and margins. For simplicity, we treat the number of components as known and present well-performing solutions for the remaining model selection problems. We adopt the expectation conditional maximum algorithm for parameter estimation. The paper is the first study in the finite mixture models literature that works with the full class of vine tree structures and a wide range of pair copula families to the best of the authors' knowledge. It combines the flexibility of vine copulas and finite mixture models to capture complex and diverse dependence structures in the multivariate data. Another contribution of the paper is a new model-based clustering algorithm, called \textit{VCMM}, that incorporates realistic interdependence structures of clusters. VCMM is interpretable and allows for various shapes of the clusters. It shows how the dependence structure varies within clusters of the data. 

The remainder of the paper is organized as follows. Section \ref{vinecop} gives an overview of vine copulas. Section \ref{vcmm} and Section \ref{vcmm-clustering} describe the vine copula mixture model and the new model-based clustering algorithm based on it. Simulation studies and analysis of real data are presented in Sections \ref{vcmm-sim} and \ref{vcmm-real}. Section \ref{conclusion} discusses our results and concludes the paper.

\section{Vine copulas}\label{vinecop}
In this section, we will recall some essential concepts of vine copulas. For more details, we refer to Chapter 3 of \cite{Joe2014}, \cite{Aas2009}, and \cite{Czado2019}.

A $d$-dimensional copula $C$ is a multivariate distribution on the unit hypercube $[0,1]^d$ with univariate uniform margins. Vine copulas are a special class of copulas based on conditioning ideas first proposed in \cite{Joe1996} and later more developed and organized by \cite{Bedford2001}. This approach allows any $d$-dimensional copula and its density to be expressed by $\frac{d\cdot(d-1)}{2}$ bivariate copulas and their densities. Moreover, the expression can be represented by an undirected graph structure \citep{Bedford2002}, i.e., \textit{regular vine (R-vine)}. A $d$-dimensional R-vine consists of $d-1$ nested trees, each tree $T_m$ has a node set $V_m$, and an edge set $E_m$ for $m=1, \ldots, d-1$. The edges in tree level $m$ are the nodes in tree level $m+1$, and a pair of nodes in tree $T_{m+1}$ are allowed to be connected if they have a common node in tree $T_m$. A formal definition of \cite{Bedford2002} is that a structure $\mathcal{V} =(T_1, \ldots, T_{d-1})$ is a regular vine on $d$ elements if it meets the following conditions:
\begin{enumerate}
\item $T_1$ is a tree with node set $V_1 = \{1, \ldots, d\}$ and edge set $E_1$.
\item $T_m$ is a tree with node set $V_m = E_{m-1}$ for $m =2, \ldots, d-1$.
\item \textit{(Proximity)} If an edge connects two nodes in $T_{m+1}$, their associated edges in $T_m$ must have a shared node in $T_m$.
\end{enumerate}
The structure $\mathcal{V} =(T_1, \ldots, T_{d-1})$ is also called a \textit{vine tree structure}, and truncating a vine after the first tree is equivalent to a Markov tree model \citep{Brechmann2012}. Thus, vines are the extending Markov trees in the sense that they allow conditional dependencies. For instance, selecting the parameters of the edges in the first tree level as correlations and the parameters of the edges in the next tree levels as partial correlations, where the partial correlations are conditioned on $m-1$ variables in tree level $m$, in a vine can represent a multivariate Gaussian distribution when margins follow a univariate normal distribution.

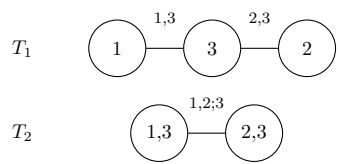
\begin{figure}[h]
	\centering
	\renewcommand{\xshiftNodes}{0.01*\linewidth}
	\renewcommand{\yshiftLabels}{.0cm}  
\begin{tikzpicture}	[every node/.style = VineNode, node distance =1.5cm,scale=0.75, transform shape]
\node[TreeLabels] (T1)       {$T_1$}
node             (v1)         [right of = T1, xshift= \xshiftNodes] {1}			
node             (v2)         [right of = v1, xshift = \xshiftNodes] {3}
node             (v3)         [right of = v2, xshift = \xshiftNodes] {2};
\draw[color=black] (v1) to node[draw=none,  font = \labelsize,fill = none, above, yshift = \yshiftLabels] {1,3} (v2);
\draw[color=black] (v2) to node[draw=none,  font = \labelsize,fill = none, above, yshift = \yshiftLabels] {2,3} (v3);
\node[TreeLabels] (T2)      [below of = T1] {$T_2$}
node             (v13)         [right of = T2, xshift =0.9cm] {1,3}
node             (v23)         [right of = v13, xshift =\xshiftNodes] {2,3};
\draw[color=black] (v13) to node[draw=none,  font = \labelsize,fill = none, above, yshift = \yshiftLabels] {1,2;3} (v23);
\end{tikzpicture}
\caption{Example of a 3-dimensional regular vine.}
\label{fig:3d-vine}
\end{figure}

Following the notation in Chapter 5 of \cite{Czado2019}, Figure \ref{fig:3d-vine} shows a $3$-dimensional vine. The vine tree structure $\mathcal{V}$ has two tree levels $T_1$ and $T_2$. The first tree level $T_1$ consists of the node set $V_1 = \{1, 2, 3\} $ and the edge set $E_1 =\{\{1, 3\}, \{2,3\}\}$. In the second tree level, the node set is $V_2 = E_1$, and the edge set is $E_2 =\{\{1,2;3\}\}$. The two nodes in $T_2$ are joined since they share the common node $3$ in $T_1$. 

To have a vine copula or \textit{pair copula construction}, each edge in a vine tree structure is associated with a bivariate copula. For example, when the bivariate copulas $C_{1,3}, C_{2,3}$ are associated with $E_1$, and the bivariate copula $C_{1,3;2}$ belongs to $E_2$ in the vine tree structure in Figure \ref{fig:3d-vine}, a vine copula model is formed. Assume each bivariate copula $C_{1,3}, C_{2,3}, C_{1,3;2}$ is parametric with the corresponding parameters $\bm{\theta}_{1,3}, \bm{\theta}_{2,3}, \bm{\theta}_{1,3;2}$ and admits the corresponding density $c_{1,3}, c_{2,3}, c_{1,3;2}$. Then we can write the density of this vine copula model for $\bm{U} = (U_1, U_2, U_3)^\top \in [0, 1]^3$ at $\bm{u} = (u_1, u_2, u_3)^\top$ as follows:
\begin{example}{The density of a vine copula model corresponding to the vine tree structure in Figure \ref{fig:3d-vine}}
\begin{equation}
\begin{aligned}
g(u_1, u_2, u_3;\bm{\theta}) =& c_{1,3}(u_1, u_3;\bm{\theta}_{1,3}) \cdot c_{2,3}(u_2, u_3;\bm{\theta}_{2,3})
\\& \cdot c_{1,2;3}\big(G_{1|3}(u_1|u_3; \bm{\theta}_{1,3}),  G_{2|3}(u_2|u_3; \bm{\theta}_{2,3});
u_3, \boldsymbol{\theta}_{1,2;3}\big),
\end{aligned} 
\label{eq:3d-density-vine}
\end{equation}
where $\bm{\theta} = (\bm{\theta}_{1,3}, \bm{\theta}_{2,3}, \bm{\theta}_{1,2;3})^\top$ contains all pair copula parameters, and $G_{1|3}$ $(G_{2|3})$ is the conditional distribution of the random variable $U_1|U_3 = u_3$ $(U_2|U_3 = u_3)$.
\label{ex:3d-density-vine}
\end{example}
Vine copulas are flexible and powerful by using bivariate building blocks (pair copulas). Moreover, Sklar's Theorem \citep{Sklar1959} states another flexibility of copulas: the density of a $d$-dimensional distribution can be decomposed into the product of its univariate marginal densities and the associated copula density. In the remainder, assume all random variables to be absolutely continuous. Then for a $d$-dimensional random vector $\bm{X} =(X_1, \ldots, X_d)^\top \in \mathbb{R}^d$ following a joint distribution $F$ with the univariate marginal distributions $F_1, \ldots F_d$ and densities $f_1, \ldots, f_d$, and a copula density $c$ of the random vector $\bm{F}=\big(F_1(X_1), \ldots F_d(X_D)\big)^\top \in [0,1]^d$, the $d$ dimensional joint density $g$ can be written as
\begin{equation}
g(\bm{x}) = c\big(F_1(x_1), \ldots, F_d(x_d)\big)\cdot f_1(x_1) \cdots f_d(x_d), \quad \bm{x} \in \mathbb{R}^d.
\label{eq:Sklar}
\end{equation} 
The vine copula model in Example \ref{ex:3d-density-vine} can formulate a $3$-dimensional joint density $g$ by Sklar's Theorem. Let $F_1, F_2, F_3$ denote parametric, univariate marginal distributions with the corresponding parameters $ \bm{\gamma}_{1},  \bm{\gamma}_{2}, \bm{\gamma}_{3}$, and $f_1, f_2, f_3$ indicate their densities.
\begin{example}{The 3-dimensional joint density with the vine copula model of Example \ref{ex:3d-density-vine} is given by}
\begin{equation}
\begin{aligned}
g(x_1, x_2, x_3;\bm{\psi}) =& c_{1,3}\big(F_{1}(x_1; \bm{\gamma}_{1}), F_{3}(x_3;\bm{\gamma}_{3});\bm{\theta}_{1,3}\big) \cdot c_{2,3}\big(F_{2}(x_2; \bm{\gamma}_{2}), F_{3}(x_3; \bm{\gamma}_{3});\bm{\theta}_{2,3}\big)
\\& \cdot c_{1,2;3}\big(F_{1|3}(x_1|x_3; \bm{\gamma}_{1}, \bm{\gamma}_{3}, \bm{\theta}_{1,3}),  F_{2|3}(x_2|x_3; \bm{\gamma}_{2}, \bm{\gamma}_{3}, \bm{\theta}_{2,3});
x_3, \boldsymbol{\theta}_{1,2;3}\big) 
\\& \cdot f_{1}(x_{1};\bm{\gamma}_{1}) \cdot f_{2}(x_{2};\bm{\gamma}_{2}) \cdot f_{3}(x_{3};\bm{\gamma}_{3}),
\end{aligned} 
\label{eq:3d-density}
\end{equation}
where the vector $\bm{\psi}$ contains the marginal and pair copula parameters, $c_{1,2;3}\big(F_{1|3}(x_1|x_3), F_{2|3}(x_2|x_3); x_3 \big)$ is the joint (copula) density corresponding to the random vector $\big(F_{1|3}(X_1|X_3), F_{2|3}(X_2 | X_3)\big)^\top$ given $X_3 = x_3$.  
\label{ex:3d-density}
\end{example}
In Example \ref{ex:3d-density}, it is assumed that all marginal distributions are parametric, which we assume in the following, and estimate the margins accordingly. Thus, we follow the \textit{inference for margins} (IFM) approach \citep{Joe1996a}. Observe that the copula density $c_{1,2;3}$ depends on the specific value $u_3$ of the conditioning variable $U_3$ in Example \ref{ex:3d-density-vine} and the specific value $x_3$ of the conditioning variable $X_3$ in Example \ref{ex:3d-density}. However, we will ignore this dependence to reduce model complexity, i.e., make the \textit{simplifying assumption}. Under the simplifying assumption, the copula density $c_{1,2;3}$ does not have any conditional dependence on the specific value of $u_3$ or $x_3$ and is a $2$-dimensional copula density. Nevertheless, it still depends on the conditioning value through its arguments. For more details, we refer to \cite{Stober2013}. 
\begin{example}{The $3$-dimensional joint density $g$ of Example \ref{ex:3d-density} under the simplifying assumption is}
\begin{equation}
\begin{aligned}
g(x_1, x_2, x_3;\bm{\psi}) =& c_{1,3}\big(F_{1}(x_1; \bm{\gamma}_{1}), F_{3}(x_3;\bm{\gamma}_{3});\bm{\theta}_{1,3}\big) \cdot c_{2,3}\big(F_{2}(x_2; \bm{\gamma}_{2}), F_{3}(x_3; \bm{\gamma}_{3});\bm{\theta}_{2,3}\big)
\\& \cdot c_{1,2;3}\big(F_{1|3}(x_1|x_3; \bm{\gamma}_{1}, \bm{\gamma}_{3}, \bm{\theta}_{1,3}),  F_{2|3}(x_2|x_3; \bm{\gamma}_{2}, \bm{\gamma}_{3}, \bm{\theta}_{2,3}); \boldsymbol{\theta}_{1,2;3}\big) 
\\& \cdot f_{1}(x_{1};\bm{\gamma}_{1}) \cdot f_{2}(x_{2};\bm{\gamma}_{2}) \cdot f_{3}(x_{3};\bm{\gamma}_{3}).
\end{aligned} 
\label{eq:3d-density-simplified}
\end{equation}
\label{ex:3d-density-simplified}
\end{example}
For a general dimension $d$, a $d$-dimensional joint density is similarly constructed using $d$ marginal densities and $\frac{d\cdot(d-1)}{2}$ associated pair-copula densities. Let $c_{e_a, e_b;D_e}$ be a parametric pair copula density associated with an edge $e$ in vine tree structure $\mathcal{V}$ and $\bm{\theta}_{e_a, e_b;D_e}$ be its parameters ($e \in E_m,\enskip \textrm{for} \enskip m=1, \ldots ,d-1$). Let $f_p$ denote a parametric, univariate marginal density with the parameters $\bm{\gamma}_p$ for $p=1, \ldots, d$. Then a $d$-dimensional joint density $g$ under the simplifying assumption can be constructed as follows:
\begin{equation}
\begin{aligned}
g(\bm{x};\bm{\psi})= &  \prod_{m=1}^{d-1} \prod_{e \in E_{m}} c_{e_a, e_b;D_e}\big( F_{{e_a} | D_e}(x_{e_a} | \bm{x}_{D_e}; \bm{\gamma}_{{e_a} | D_e},\boldsymbol{\theta}_{{e_a} | D_e}), F_{{e_b} | D_e}\bigl(x_{e_b} | \bm{x}_{D_e}; \bm{\gamma}_{{e_b} | D_e}, \bm{\theta}_{{e_b} | D_e}); \bm{\theta}_{e_a, e_b;D_e}\big)
\\ & \cdot\prod_{p=1}^{d} f_{p}(x_{p};\bm{\gamma}_{p}),
\end{aligned}
\label{eq:joint-density-d-simplified}
\end{equation}
where $\bm{x}_{D_e} = (x_z)_{z \in D_e} $ is a subvector of $\bm{x} = (x_1, \ldots, x_d)^\top \in \R^d$,  the parameter vector $\bm{\psi}$ contains the marginal and pair copula parameters, $F_{{e_a} | D_e}$ is the conditional distribution function of the random variable $X_{e_a}|\bm{X}_{D_e} = \bm{x}_{D_e}$. It can be calculated recursively \citep{Joe1996}. The marginal $\bm{\gamma}_{{e_a} | D_e}$ and pair copula $\boldsymbol{\theta}_{{e_a} | D_e}$ parameters are used to determine $F_{{e_a} | D_e}$. The set $D_e$ is called the \textit{conditioning set}, and the indices $e_a, e_b$ form the \textit{conditioned set}. $D_e$ has $m-1$ elements in tree level $m$; therefore, it is empty in the first tree.
\section{Vine copula mixture models}\label{vcmm}
We now introduce a vine copula mixture model and describe approaches for model selection and parameter estimation problems in the following subsections. The model is fully parametric, i.e., it works with parametric pair copulas and univariate marginal distributions.
\subsection{Model formulation} \label{vcmm-formulation}
Suppose the data consists of $n$ observations, where an observation $\bm{x}_i = (x_{i,1}, \ldots, x_{i,d})^\top$ is an independent realization of a $d$-dimensional random vector $\bm{X} = (X_1, \ldots, X_d)^\top$ for $i=1, \ldots, n$. Assume that a mixture of $k$ components ($k \in \R^{+}$) generates the data, and the density $g_{j}$ of the $j$th component for $j=1, \ldots, k$ can be stated by Sklar's Theorem as given in Equation (\ref{eq:joint-density-d-simplified}). Assume that, additionally, the $j$th component has a mixture weight $\pi_j$ with $\pi_j \in (0,1)$ for $j=1, \ldots, k$ and $\sum\limits_{j}^{k} \pi_j =1$. Then the density of the vine copula mixture model for $\bm{X} = (X_1, \ldots, X_d)^\top$ at $\bm{x} = (x_1, \ldots, x_d)^\top$ can be written as:
\begin{equation}
g(\bm{x};\bm{\eta}) = \sum_{j=1}^{k} \pi_j \cdot g_{j}(\bm{x};\bm{\psi}_j). \label{eq:vcmm-fmm}
\end{equation}
Here the vector $\bm{\psi}_j$ contains the marginal and pair copula parameters of the $j$th component, $\bm{\eta}$ denotes all model parameters, i.e., $\bm{\eta} =(\bm{\eta}_1, \ldots, \bm{\eta}_k)^\top$ and $\bm{\eta}_j = (\pi_j, \bm{\psi}_j)^\top$ for $j=1, \ldots, k$.
\begin{example} {Vine copula mixture model formulation in three dimensions with two components}

Assume data, where an observation $\boldsymbol{x}_i =\big(x_{i,1}, x_{i,2}, x_{i,3}\big)^\top$ for $i = 1, \ldots, n$ is given, and there are two components generating the data with the mixture weights $\pi_1>0$ and $\pi_2>0$ ($\pi_1  + \pi_2 =1$). An observation of the first and second component is an independent realization of a $3$-dimensional random vector $\bm{X}_{(1)} = (X_{1(1)}, X_{2(1)}, X_{3(1)})^\top$ and  $\bm{X}_{(2)} = (X_{1(2)}, X_{2(2)}, X_{3(2)})^\top$, respectively. Figure \ref{fig:vcmm-form-1} shows the vine copula model of each component. $T_{(1)1}$ and $T_{(2)1}$ represent the first tree, whereas $T_{(1)2}$ and $T_{(2)2}$ refer to the second tree of the first and second component. The edge sets of the first component are $E_{(1)1} =\{\{1,2\}, \{2,3\}\}$, $E_{(1)2} =\{1,3;2\}$ and that of the second component are $E_{(2)1} =\{\{1,2\}, \{1,3\}\}$, $E_{(2)2} =\{2,3;1\}$, respectively.
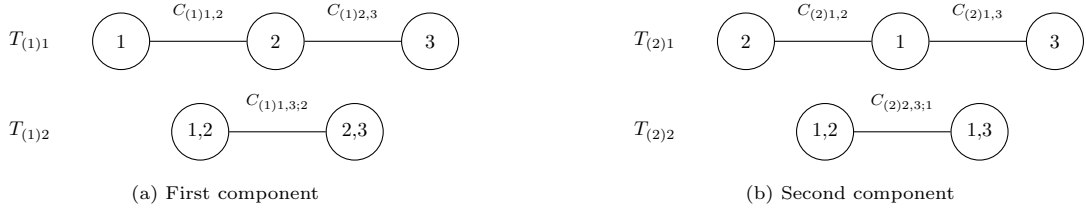
\begin{figure}[H]
\begin{subfigure}[b]{0.45\linewidth}
	\centering
	\renewcommand{\xshiftNodes}{0.15*\linewidth}
	\renewcommand{\yshiftLabels}{.0cm}  
\begin{tikzpicture}	[every node/.style = VineNode, node distance =1.6cm,scale=0.75, transform shape]]
\node[TreeLabels] (T1) {$T_{(1)1}$}
node             (v1)         [right of = T1] {1}			
node             (v2)         [right of = v1, xshift = \xshiftNodes] {2}
node             (v3)         [right of = v2, xshift = \xshiftNodes] {3};
\draw[color=black] (v1) to node[draw=none, font = \labelsize, fill = none, above, yshift = \yshiftLabels] {$C_{(1)1,2}$} (v2);
\draw[color=black] (v2) to node[draw=none,  font = \labelsize,fill = none, above, yshift = \yshiftLabels] {$C_{(1)2,3}$} (v3);
\node[TreeLabels] (T2)      [below of = T1] {$T_{(1)2}$}
node             (v13)         [right of = T2, xshift = 1.25*\xshiftNodes] {1,2}
node             (v23)         [right of = v13, xshift =\xshiftNodes] {2,3};
\draw[color=black] (v13) to node[draw=none, font = \labelsize, fill = none, above, yshift = \yshiftLabels] {$C_{(1)1,3;2}$} (v23);
\end{tikzpicture}
\caption{First component}
\end{subfigure}
\qquad
\begin{subfigure}[b]{0.45\linewidth}
	\centering
	\renewcommand{\xshiftNodes}{0.15*\linewidth}
	\renewcommand{\yshiftLabels}{.0cm}  
\begin{tikzpicture}	[every node/.style = VineNode, node distance =1.6cm,scale=0.75, transform shape]
\node[TreeLabels] (T1)    {$T_{(2)1}$}
node             (v1)         [right of = T1] {2}			
node             (v2)         [right of = v1, xshift = \xshiftNodes] {1}
node             (v3)         [right of = v2, xshift =\xshiftNodes] {3};
\draw[color=black] (v1) to node[draw=none,  font = \labelsize,fill = none, above, yshift = \yshiftLabels] {$C_{(2)1,2}$} (v2);
\draw[color=black] (v2) to node[draw=none, font = \labelsize, fill = none, above, yshift = \yshiftLabels] {$C_{(2)1,3}$} (v3);
\node[TreeLabels] (T2)      [below of = T1] {$T_{(2)2}$}
node             (v13)         [right of = T2, xshift = 1.25*\xshiftNodes] {1,2}
node             (v23)         [right of = v13, xshift =\xshiftNodes] {1,3};
\draw[color=black] (v13) to node[draw=none,  font = \labelsize,fill = none, above, yshift = \yshiftLabels] {$C_{(2)2,3;1}$} (v23);
\end{tikzpicture}
\caption{Second component}
\end{subfigure}
\caption{Vine copula model of two components.}
\label{fig:vcmm-form-1}
\end{figure}
The pair copula families and marginal distributions of both components are parametric. $C_{(1)1,2}$, $C_{(1)2,3}$, $C_{(1)1,3;2}$ denote pair copula families of the first component associated with the edges in $E_{1(1)}$ and $E_{2(1)}$, while $\bm{\theta}_{(1)1,2}$, $\bm{\theta}_{(1)2,3}$, $\bm{\theta}_{(1)1,3;2}$ show the corresponding parameters. For the second component, we use notations $C_{(2)1,2}$, $C_{(2)1,3}$, $C_{(2)2,3;1}$ for pair copula families associated with the edges in $E_{1(2)}$ and $E_{2(2)}$, and $\bm{\theta}_{(2)1,2}$, $\bm{\theta}_{(2)1,3}$, $\bm{\theta}_{(2)2,3;1}$ refer to the corresponding parameters. A copula density is denoted by the small $c$ letter. $F_{1(1/2)}, F_{2(1/2)}, F_{3(1/2)}$ refer to the marginal distributions of the random variables $X_{1(1/2)}$, $X_{2(1/2)}$, $X_{3(1/2)}$ with the corresponding parameters $\bm{\gamma}_{1(1/2)}, \bm{\gamma}_{2(1/2)}, \bm{\gamma}_{3(1/2)}$ for the first/second component. The small $f$ letter denotes a marginal density. We can write the \textbf{density of the first component} at $\bm{x} =(x_1, x_2, x_3)^\top$:
\begin{equation}
\begin{aligned}
g_{1}(\bm{x};\bm{\psi}_1) =& c_{(1)1,2}\big(F_{1(1)}(x_1; \bm{\gamma}_{1(1)}), F_{2(1)}(x_2;\bm{\gamma}_{2(1)});\bm{\theta}_{(1)1,2}\big) \cdot c_{(1)2,3}\big(F_{2(1)}(x_2; \bm{\gamma}_{2(1)}), F_{3(1)}(x_3; \bm{\gamma}_{3(1)});\bm{\theta}_{(1)2,3}\big)
\\& \cdot c_{(1)1,3;2}\big(F_{(1)1|2}(x_1|x_2; \bm{\gamma}_{1(1)}, \bm{\gamma}_{2(1)}, \bm{\theta}_{(1)1,2}),  F_{(1)3|2}(x_3|x_2; \bm{\gamma}_{3(1)}, \bm{\gamma}_{2(1)}, \bm{\theta}_{(1)2,3}); \boldsymbol{\theta}_{(1)1,3;2}\big) 
\\& \cdot f_{1(1)}(x_{1};\bm{\gamma}_{1(1)}) \cdot f_{2(1)}(x_{2};\bm{\gamma}_{2(1)}) \cdot f_{3(1)}(x_{3};\bm{\gamma}_{3(1)}),
\end{aligned} 
\label{eq:vcmm-ex-comp-1}
\end{equation}
where the pair copula parameters used to determine the conditional distribution functions $F_{(1)1|2}$ and $F_{(1)3|2}$ are given by $\bm{\theta}_{(1)1|2} = \bm{\theta}_{(1)1,2}$ and $\bm{\theta}_{(1)3|2} = \bm{\theta}_{(1)2,3}$, respectively. The marginal parameters needed for the same calculation are denoted by $\bm{\gamma}_{(1)1|2} = (\bm{\gamma}_{1(1)}, \bm{\gamma}_{2(1)})^\top$ and $\bm{\gamma}_{(1)3|2} = (\bm{\gamma}_{2(1)}, \bm{\gamma}_{3(1)})^\top$. We show the marginal and pair copula parameters of the first component by $\bm{\psi}_1 = (\bm{\gamma}_1, \bm{\theta}_1)^\top$, where $\bm{\gamma}_{1} =(\bm{\gamma}_{1(1)}, \bm{\gamma}_{2(1)},\bm{\gamma}_{3(1)})^\top$ and $\bm{\theta}_1 = (\bm{\theta}_{(1)1,2}, \bm{\theta}_{(1)2,3}, \bm{\theta}_{(1)1,3;2})^\top$.
As in the first component, we can define the parameters and write the \textbf{second component density} at $\bm{x} =(x_1, x_2, x_3)^\top$:
\begin{equation}
\begin{aligned}
g_{2}(\bm{x};\bm{\psi}_2) =& c_{(2)1,2}\big(F_{1(2)}(x_1; \bm{\gamma}_{1(2)}), F_{2(2)}(x_2;\bm{\gamma}_{2(2)});\bm{\theta}_{(2)1,2}\big) \cdot c_{(2)1,3}\big(F_{1(2)}(x_1; \bm{\gamma}_{1(2)}), F_{3(2)}(x_3; \bm{\gamma}_{3(2)});\bm{\theta}_{(2)1,3}\big)
\\& \cdot c_{(2)2,3;1}\big(F_{(2)2|1}(x_2|x_1; \bm{\gamma}_{2(2)}, \bm{\gamma}_{1(2)}, \bm{\theta}_{(2)1,2}),  F_{(2)3|1}(x_3|x_1; \bm{\gamma}_{3(2)}, \bm{\gamma}_{1(2)}, \bm{\theta}_{(2)1,3}); \boldsymbol{\theta}_{(2)2,3;1}\big) 
\\& \cdot f_{1(2)}(x_{1};\bm{\gamma}_{1(2)}) \cdot f_{2(2)}(x_{2};\bm{\gamma}_{2(2)}) \cdot f_{3(2)}(x_{3};\bm{\gamma}_{3(2)}),
\end{aligned} 
\label{eq:vcmm-ex-comp-2}
\end{equation}
As a result, the \textbf{vine copula mixture model density} at $\bm{x} =(x_1, x_2, x_3)^\top$ is given by
\begin{equation}
g(\bm{x};\bm{\eta}) = \pi_{1}g_1(\bm{x};\bm{\psi}_{1}) + \pi_{2}g_2(\bm{x};\bm{\psi}_{2}), \label{eq:vcmm-ex-all}
\end{equation}
where $\bm{\eta}_1 = (\pi_1, \bm{\psi}_1)^\top$ and $\bm{\eta}_2 = (\pi_2, \bm{\psi}_2)^\top$ indicate the model parameters of the first and second component. All model parameters are given by $\bm{\eta} = (\bm{\eta}_1, \bm{\eta}_2)^\top$.
\end{example}
\subsection{Model selection} \label{vcmm-selection}
Vine copula mixture models inherit the problem of estimating the total number of components $k$ hidden in the data from the finite mixture models. Moreover, due to its formulation in Equations (\ref{eq:joint-density-d-simplified}) and (\ref{eq:vcmm-fmm}), the vine tree structure $\mathcal{V}_j$, pair copula families $\mathcal{B}_j(\mathcal{V}_j)$, and marginal distributions $\mathcal{F}_j = \{F_{1(j)}, \ldots, F_{d(j)}\}$ of the $j$th component need to be chosen for $j=1, \ldots, k$. Accordingly, pair copula parameters $\bm{\theta}_j(\mathcal{B}_j(\mathcal{V}_j))$ and marginal parameters $\bm{\gamma}_j(\mathcal{F}_j )$ should be estimated for $j=1, \ldots, k$. To simplify, we will assume that the total number of components generating the data is known. If a priori information about $k$ does not exist, methods that estimate it from the data need to be developed for vine copula mixture models. However, it is currently not the focus of our work. Instead, we will explain the approaches to the remaining model selection problems in the following and discuss the parameter estimation in Section \ref{vcmm-param-est}.

Assume an observation $\bm{x}_i = (x_{i,1}, \ldots, x_{i,d})^\top$ is assigned to a component for $i=1, \ldots, n$. We will learn model components with the assignment, use the vine copula mixture models for clustering applications, and explain assignment approaches in Algorithm \ref{vcmm-clustering-algo} in Section \ref{vcmm-clustering}. We will show the promising results of this approach in Sections \ref{vcmm-sim} and \ref{vcmm-real} and discuss a modification, when it is hard to learn model components, e.g., components have non-negligible overlaps, in Section \ref{vcmm-sim-4}.

Assume the total number of observations assigned to the $j$th component is $n_j$, and the observations belonging to the $j$th component are given by $\bm{x}_{(j)i_j} = (x_{(j)i_j,1}, \ldots, x_{(j)i_j,d})^\top$ for $i_j = 1, \ldots n_j$ and $j=1, \ldots, k$. It holds that $\sum\limits_{j=1}^{k} n_j = n$ and $\underset{\forall (j, i_j)}{\bigcup}\, \bm{x}_{(j)i_j} = \underset{\forall i}{\bigcup}\, \bm{x}_i $. We denote the $p$th variable in the $j$th component by $\bm{x}_{p(j)} = (x_{(j)1,p}, \ldots, x_{(j)n_j,p})^\top$ for $p = 1, \ldots d$ and $j=1, \ldots, k$. With the given assignment, we first select the marginal distributions of each cluster. Such a candidate set could be prespecified or chosen by a data analysis such as a histogram and quantile-quantile (QQ) plot. Then the marginal distribution family $F_{p(j)}$ for the variable $\bm{x}_{p(j)}$ will be determined using a model selection criteria. More precisely: 

\begin{itemize}
\item \textbf{\textit{Marginal distribution selection}} $\mathcal{F}_j$: For $p=1, \ldots, d$ and $j=1, \ldots, k$, the log-likelihood of the marginal distribution $F_{p(j)}$ with the density $f_{p(j)}$ and parameters $\bm{\gamma}_{p(j)}$ on the variable $\bm{x}_{p(j)}$ is
\begin{equation}
\ell(\bm{\gamma}_{p(j)}) = \sum\limits_{i=1}^{n_j} \textrm{log } \big(f_{p(j)} (x_{(j)i,p}; \bm{\gamma}_{p(j)})\big) \quad \textrm{for} \quad p = 1, \ldots d \quad  \textrm{and} \quad j = 1, \ldots k. 
\label{eq:loglik-margin}
\end{equation}
The BIC, a commonly used model selection criteria, is given by
\begin{equation}
BIC(\bm{\gamma}_{p(j)}) = -2 \cdot \ell(\bm{\gamma}_{p(j)}) +|\bm{\gamma}_{p(j)}|  \cdot \text{log }(n_j) \quad \textrm{for} \quad p = 1, \ldots d \quad  \textrm{and} \quad j = 1, \ldots k,
\label{eq:BIC1}
\end{equation}
where $|\bm{\gamma}_{p(j)}|$  refers to the number of marginal parameters in $F_{p(j)}$, and $n_j$ denotes the total number of observations in $\bm{x}_{p(j)}$. For each candidate for the marginal distribution on the variable $\bm{x}_{p(j)}$, first, the parameters that maximize the log-likelihood $\ell(\bm{\gamma}_{p(j)})$ are estimated, then the marginal distribution family with the lowest BIC is selected. 
\end{itemize}
Since the joint density can be decomposed into univariate marginal densities and a vine copula density by Sklar's Theorem, we now estimate the u-data for a vine copula model by applying the probability integral transformation using the estimated margins $\hat{F}_{p(j)}$ for each cluster: $\bm{\hat{u}}_{p(j)} = \hat{F}_{p(j)}(\bm{x}_{p(j)}; \bm{\hat{\gamma}}_{p(j)})$ and set $\bm{\hat{u}}_{p(j)} = (\hat{u}_{(j)1,p}, \ldots, \hat{u}_{(j)n_j,p})^\top$. After obtaining the u-data of the $j$th component, the best selection of $\mathcal{V}_j$ would be the true structure selection, but the total number of vine tree structures on $d$ variables is $\frac{d!}{2} \cdot 2^{{d-2 \choose 2}}$ \citep{Morales-Napoles2010}. If $d$ is small, it is possible to enumerate all possible structures. However, it is usually not a feasible approach, even with small dimensions, as one also needs to select pair copula families for each scenario. Therefore, a greedy algorithm proposed for the vine tree and pair copula family selection might be used for each component. For instance, we use the greedy algorithm of \cite{Dissmann2013} in Algorithm \ref{vcmm-clustering-algo}. Briefly, it works as follows:
\begin{itemize}
\item \textbf{\textit{Vine tree structure selection}} $\mathcal{V}_j$: For $j=1, \ldots, k$, it proceeds sequentially tree by tree, starting from tree one, and finds the maximum spanning tree at each tree among all edges allowed by proximity. Edge weight is the absolute empirical Kendall's $\tau$ value between the pair of nodes forming the edge.
\item \textbf{\textit{Pair copula family selection}} $\mathcal{B}_j(\mathcal{V}_j)$: For $j=1, \ldots, k$, after learning the vine tree structure, pair copula families of the given structure are also estimated sequentially tree by tree. For a  parametric pair copula $C_{(j)e_a, e_b;D_e}$ associated with an edge $e$ in $\mathcal{V}_j$ with the density $c_{(j)e_a, e_b;D_e}$ and parameters $\bm{\theta}_{(j)e_a, e_b;D_e}$, one first estimates the parameters that maximize the log-likelihood $\ell(\bm{\theta}_{(j)e_a, e_b;D_e})$. Later the copula family with the lowest Akaike Information Criterion (AIC) \citep{Akaike1998} is chosen:
\begin{equation}
AIC(\bm{\theta}_{(j)e_a, e_b;D_e}) = -2 \cdot \ell(\bm{\theta}_{(j)e_a, e_b;D_e})+ 2 \cdot |\bm{\theta}_{(j)e_a, e_b;D_e}| \quad \textrm{for} \quad  j = 1, \ldots k \quad \textrm{and}\quad e\in\mathcal{V}_j,
\label{eq:AIC-2}
\end{equation}
where $|\bm{\theta}_{(j)e_a, e_b;D_e}|$ denotes the number of copula parameters in $c_{(j)e_a, e_b;D_e}$. One does not need an alternative selection criterion like the BIC to induce model sparsity for the pair copula family selection when the fitted pair copula families have one or two parameters as later proposed in Algorithm \ref{vcmm-clustering-algo}.
\end{itemize}
\subsection{Parameter estimation} \label{vcmm-param-est}
Given that the marginal distributions, vine tree structure, and associated pair copula families of each component are selected and known, a first task in the vine copula mixture model is to estimate the model parameters $\bm{\eta}$ in Equation (\ref{eq:vcmm-fmm}). The optimal parameter estimates would be the values that maximize the log-likelihood of the given data defined in Equation (\ref{eq:vcmm-loglik-true}):
\begin{equation}
\ell(\boldsymbol{\eta}) = \textrm{log }  \prod\limits_{i=1}^{n} g(\bm{x}_i;\bm{\psi}) = \textrm{log }  \prod\limits_{i=1}^{n} \sum\limits_{j=1}^{k} \pi_j \cdot g_{j}(\bm{x}_i;\bm{\psi}_j).
\label{eq:vcmm-loglik-true}
\end{equation}
Nevertheless, the true assignment of the observations to each component is unknown, and the parameter estimates would change depending on the component to which an observation belongs. As a solution to this problem, the expectation-maximization (EM) algorithm \citep{Dempster1977} views the observations $\bm{x}_i = (x_{i,1}, \ldots, x_{i,d})^\top$ as incomplete and introduces latent variables  $\bm{z}_i = (z_{i,1}, \ldots, z_{i,k})^\top$, where each element $z_{i,j}$ is a binary variable defined as
\begin{equation}
z_{i,j} = 
\begin{cases} 
       & 1,  \quad  \textrm{if } \bm{x}_i\textrm{ belongs to the \textit{j}th component,} \\
       & 0,   \quad   \textrm{otherwise,}
\end{cases} 
\end{equation}
and $\sum\limits_{j=1}^{k} z_{i,j} = 1$. The random vector $\boldsymbol{Z}_i$ corresponding to $\boldsymbol{z}_i$ follows a multinomial distribution with one trial and probabilities 
$\pi_1, \ldots, \pi_k$, that is, $\boldsymbol{Z}_i \sim Mult\big(1, \boldsymbol{\pi}=(\pi_1, \ldots, \pi_k)\big)$. Then we can write the \textit{complete data log-likelihood} $\ell_{c}(\boldsymbol{\eta}; \boldsymbol{z}, \boldsymbol{x})$ of the complete data $\bm{y}_i =(\bm{x}_i, \bm{z}_i)^\top$ from Equation (\ref{eq:vcmm-fmm}) as:
\begin{equation}
\ell_{c}(\boldsymbol{\eta}; \boldsymbol{z}, \boldsymbol{x}) = \textrm{log }  \prod\limits_{i=1}^{n} \prod\limits_{j=1}^{k} [\pi_j \cdot g_{j}(\bm{x}_i;\bm{\psi}_j)]^{z_{i,j}} = \sum\limits_{i=1}^{n} \sum\limits_{j=1}^{k} z_{i,j} \cdot \textrm{log }\pi_j + \sum\limits_{i=1}^{n} \sum\limits_{j=1}^{k} z_{i,j} \cdot \textrm{log }g_{j}(\bm{x}_i;\bm{\psi}_j),
\label{eq:vcmm-loglik}
\end{equation}
where $g_{j}(\bm{x}_i;\bm{\psi}_j)$ is given in Equation (\ref{eq:joint-density-d-simplified}). Hence, we can write:
\begin{equation}
\begin{aligned}
\ell_{c}(\boldsymbol{\eta}; \boldsymbol{z}, \boldsymbol{x})  =& \sum\limits_{i=1}^{n} \sum\limits_{j=1}^{k} [z_{i,j} \cdot \textrm{log }\pi_j] + \sum\limits_{i=1}^{n} \sum\limits_{j=1}^{k} \sum\limits_{p=1}^{d} [z_{i,j} \cdot \textrm{log } f_{p(j)}(x_{i,p};\bm{\gamma}_{p(j)})] \\
& +  \sum\limits_{i=1}^{n} \sum\limits_{j=1}^{k} \sum_{m=1}^{d-1} \sum_{e \in E_{(j)m}}  \big[z_{i,j} \cdot  \text{log } c_{(j)e_a, e_b;\bm{D_e}}\big( F_{(j){e_a} | \bm{D_e}}(x_{i, e_a} | \bm{x}_{i,\bm{D_e}}; \bm{\gamma}_{(j){e_a} | \bm{D_e}},\bm{\theta}_{(j){e_a} | \bm{D_e}}), 
\\&F_{(j){e_b} | \bm{D_e}}\bigl(x_{i,e_b} | \bm{x}_{i,\bm{D_e}}; \bm{\gamma}_{(j){e_b} | \bm{D_e}}, \bm{\theta}_{(j){e_b} | \bm{D_e}});\bm{\theta}_{(j)e_a, e_b;\bm{D_e}}\big)\big].
\end{aligned}
\label{eq:vcmm-loglik-2}
\end{equation}
Note that we use $\bm{\theta}_j$ for the pair copula parameters instead of $\bm{\theta}_j(\mathcal{B}_j(\mathcal{V}_j))$ and $\bm{\gamma}_j$ for the marginal parameters instead of $\bm{\gamma}_j(\mathcal{F}_j)$ to simplify notation.

The EM algorithm alternates the E and M steps, increasing the data log-likelihood at each iteration \citep{Dempster1977}. The E-step requires calculating the conditional expectation of the complete-data log likelihood, given the observed data and current parameter estimates. The M-step maximizes the expected complete data log-likelihood from the E-step over all parameters. We need to estimate marginal parameters $\bm{\gamma}_j$, pair copula parameters $\bm{\theta}_j$, and mixture weight $\pi_j$ of the $j$th component. Since their joint estimation is not tractable and efficient, we use the expectation conditional maximum (ECM) algorithm \citep{Meng1993}. Here, the M-step in the EM is replaced by three lower dimensional maximization problems called CM-steps. The vine tree structure, associated pair copula families and marginal distributions remain fixed at the initial choice for each component in the ECM iterations. Then the $(t+1)$th iteration steps are
\begin{enumerate}
\item \textit{\textbf{E-step} (Posterior probabilities)}: Calculate the posterior probability that an observation $\bm{x}_i$ belongs to the $j$th component given the current values of the model parameters $\pi^{(t)}_j$ and $\bm{\psi}^{(t)}_j =(\bm{\gamma}^{(t)}_j, \bm{\theta}^{(t)}_j)^\top$:
\begin{equation}
r_{i,j}^{(t+1)} = \frac{\pi_j^{(t)} g_j(\bm{x}_i;\bm{\psi}_{j}^{(t)})} { \sum\limits_{j'=1}^{k} \pi_{j'}^{(t)} g_{j'}(\bm{x}_i;\bm{\psi}_{j'}^{(t)})}\quad \textrm{for} \quad i = 1, \ldots n \quad  \textrm{and} \quad j = 1, \ldots k.
\label{eq:E-step}
\end{equation}
\item \textit{\textbf{CM-step 1} (Mixture weights)}: Maximize $\ell_{c}(\boldsymbol{\eta}; \boldsymbol{z}, \boldsymbol{x})$ over $\pi_j$ given the updated posterior probabilities $r_{i,j}^{(t+1)}$:
\begin{equation}
 \pi^{(t+1)}_j = \argmax_{\pi_j}\sum\limits_{i=1}^{n} r^{(t+1)}_{i,j} \cdot \textrm{log }\pi_j  \quad \textrm{for} \quad j = 1, \ldots k.
\label{eq:CM-step-11}
\end{equation}
A closed form solution exists and is given by
\begin{equation}
\pi_j^{(t+1)} =\frac{\sum\limits_{i=1}^{n}r_{i,j}^{(t+1)}}{n} \quad \textrm{for} \quad j = 1, \ldots k.
\label{eq:CM-step-1}
\end{equation}
\item \textit{\textbf{CM-step 2} (Marginal parameters)}: Optimal marginal parameter estimates $\bm{\gamma}^{*}_j$ of the $j$th component would maximize $\ell_{c}(\boldsymbol{\eta}; \boldsymbol{z}, \boldsymbol{x})$ over $\bm{\gamma}_j$ given the current values of the pair copula parameters $\bm{\theta}^{(t)}_j$ and updated posterior probabilities $r_{i,j}^{(t+1)}$:
\begin{equation}
\bm{\gamma}^{*}_j= \argmax_{\bm{\gamma}_j} \sum\limits_{i=1}^{n} r_{i,j}^{(t+1)} \cdot \textrm{log }g_{j}(\bm{x}_i;\bm{\gamma}_j, \bm{\theta}^{(t)}_j) \quad \textrm{for} \quad j = 1, \ldots k.
\label{eq:CM-step-21}
\end{equation}
However, since a closed form solution does not exist, we numerically maximize $\ell_{c}(\boldsymbol{\eta}; \boldsymbol{z}, \boldsymbol{x})$ over $\bm{\gamma}_j$
\begin{equation}
\max_{\bm{\gamma}_j} \sum\limits_{i=1}^{n}  r_{i,j}^{(t+1)} \cdot \textrm{log }g_{j}(\bm{x}_i;\bm{\gamma}_j, \bm{\theta}^{(t)}_j) \quad \textrm{for} \quad j = 1, \ldots k
\label{eq:CM-step-2}
\end{equation}
to find the updated values of the marginal parameters $\bm{\gamma}^{(t+1)}_j$.
\item \textit{\textbf{CM-step 3} (Pair copula parameters)}: Again, a closed form solution that maximizes $\ell_{c}(\boldsymbol{\eta}; \boldsymbol{z}, \boldsymbol{x})$ over $\bm{\theta}_j$ given the current values of the marginal parameters $\bm{\gamma}^{(t+1)}_j$ and  updated posterior probabilities $r_{i,j}^{(t+1)}$ does not exist. Hence, we numerically maximize $\ell_{c}(\boldsymbol{\eta}; \boldsymbol{z}, \boldsymbol{x})$ over $\bm{\theta}_j$
\begin{equation}
\max_{\bm{\theta}_j} \sum\limits_{i=1}^{n} r_{i,j}^{(t+1)} \cdot \textrm{log }g_{j}(\bm{x}_i;\bm{\gamma}^{(t+1)}_j, \bm{\theta}_j) \quad \textrm{for} \quad j = 1, \ldots k
\label{eq:CM-step-3}
\end{equation}
to find the updated values of the pair copula parameters $\bm{\theta}^{(t+1)}_j$. In the case of a $d$-dimensional vine copula with parametric pair copulas, the total number of pair copula parameters to be estimated grows quadratically in dimension $d$. However, truncating a vine tree structure at tree level 1, i.e., obtaining a Markov tree, reduces the total number of parameter estimates linear in dimension $d$, one of the main motivations for the third step's formulation with Markov trees in Algorithm \ref{vcmm-clustering-algo}. 
\end{enumerate}
\textit{\textbf{Starting values and stopping condition}}

For the ECM algorithm, we require initial parameters $\bm{\eta}^{(0)}_j$ of the $j$th component for $j=1, \ldots, k$. Since the marginal distributions, vine tree structure, and associated pair copula families of each component stay fixed in the ECM iterations, we, additionally, need to select them. One method is to select an initial partition in advance. A more general alternative, given in Algorithm \ref{vcmm-clustering-algo}, uses quick clustering algorithms with possible weights of observations $\bm{x}_i = (x_{i,1}, \ldots, x_{i,d})^\top$ for $i=1, \ldots, n$ in different components to have a starting partition. Assume that the total number of observations assigned to the $j$th component at the $0$th iteration is $n^{(0)}_j$, and the observations belonging to the $j$th component at the $0$th iteration are given by $\bm{x}^{(0)}_{(j)i_j} = (x^{(0)}_{(j)i_j,1}, \ldots, x^{(0)}_{(j)i_j,d})^\top$ for $i_j = 1, \ldots, n^{(0)}_j$ and $j=1, \ldots, k$. It holds that $\sum\limits_{j=1}^{k} n^{(0)}_j = n$ and $\underset{\forall (j, i_j)}{\bigcup}\, \bm{x}^{(0)}_{(j)i_j} = \underset{\forall i}{\bigcup}\, \bm{x}_i $. After specifying an initial set for parametric marginal distributions, vine tree structures and associated parametric pair copula families, the starting values can be obtained as follows:
\begin{enumerate}
\item Initial marginal distributions $F^{(0)}_{p(j)}$ and marginal parameters $\bm{\gamma}^{(0)}_{p(j)}$: For $j = 1, \ldots, k$ and $p = 1, \ldots, d$, the marginal parameters maximize the log-likelihood of a variable $\bm{x}^{(0)}_{p(j)} = (x^{(0)}_{(j)1,p}, \ldots, x^{(0)}_{(j)n^{(0)}_j,p})^\top$ given in Equation (\ref{eq:loglik-margin}), then the marginal distribution with the lowest BIC given in Equation (\ref{eq:BIC1}) is chosen as described in Section \ref{vcmm-selection}. 
\item Initial vine tree structure $\mathcal{V}^{(0)}_j$, pair copula families $\mathcal{V}^{(0)}_j(\mathcal{B}^{(0)}_j)$ and its parameters $\bm{\theta}^{(0)}_{(j)}$: For $j = 1, \ldots, k$ and $p = 1, \ldots, d$, first, the cumulative distribution function of the chosen marginal distribution $F^{(0)}_{p(j)}$ with parameters $\bm{\gamma}^{(0)}_{p(j)}$ on the variable $\bm{x}^{(0)}_{p(j)}$ is fitted to obtain u-data of the $j$th component, $\bm{u}^{(0)}_{p(j)} = F^{(0)}_{p(j)}(\bm{x}^{(0)}_{p(j)}; \bm{\gamma}^{(0)}_{p(j)})$. The estimated u-data is then used to select a vine or Markov tree and the associated pair copula families with their parameters as discussed in Section \ref{vcmm-selection}.
\item Initial mixture weights $\pi^{(0)}_{(j)}$: For $j = 1, \ldots, k$, they are proportional to the total number of observations belonging to the $j$th component and given by $\pi^{(0)}_{(j)} = \frac{n^{(0)}_j}{n}$.
\end{enumerate}
The ECM is sensitive to the starting values, and a poor choice of them may result in convergence to a local maximum as an EM-type algorithm \citep{Karlis2003}. However, our starting values optimize an associated model selection criterion based on the data log-likelihood. Even though there is no guarantee that the vine copula mixture model initialized by a fast clustering algorithm will select the correct model components, and its ECM iterations will converge to the global optimum, we show promising results of our current setup and discuss modifications with simulation studies in Section \ref{vcmm-sim} and real data sets in Section \ref{vcmm-real}. 

A stopping criterion terminates the algorithm when the relative difference in the data log-likelihood between two successive iterations is less than the desired tolerance. We use the tolerance level ($tol$) 0.00001 in our simulation studies and real data analysis for clustering.
\begin{equation}
\frac{\ell(\bm{\eta}^{(t+1)}) - \ell(\bm{\eta}^{(t)})}{\ell(\bm{\eta}^{(t)})} <   tol\quad  \textrm{for} \quad t = 1, \ldots, (s-1). 
\label{eq:stopping}
\end{equation}
\section{Model-based clustering algorithm: VCMM}\label{vcmm-clustering}
This section will formulate a model-based clustering algorithm (VCMM) with vine copula mixture models introduced in Section \ref{vcmm}. We provide our pseudo-code in Algorithm \ref{vcmm-clustering-algo} and give an overview flowchart in \ref{app-flow}. It consists of six primary building blocks and implements them in {\fontfamily{pcr}\selectfont R} \citep{R}. 

The first step (initial clustering assignment via a fast clustering algorithm) partitions observations $\bm{x}_i = (x_{i,1}, \ldots, x_{i,d})^\top$ for $i=1, \ldots, n$ into $k$ components (clusters) using a quick clustering algorithm. Our analyses will mostly use the distance-based clustering, k-means \citep{Hartigan1979} algorithm, with default specifications used in the package {\fontfamily{pcr}\selectfont stats}. Alternative clustering algorithms or partitions could be specified based on the analyzed data set. We will present an exemplary scenario in Section \ref{vcmm-sim-4}. For clustering algorithms, which are sensitive to the variables' scale, like k-means, we apply a standardization for each variable using the function {\fontfamily{pcr}\selectfont scale}. Since these algorithms might depend on the seed, we will present a real data analysis in Section \ref{AIS} to guide choosing a good option.

As a second step, we select an initial VCMM model. The marginal distributions are determined among a candidate set of univariate parametric distributions by a model selection criteria. As discussed in Section \ref{vcmm-selection}, the parametric families in the candidate set could be chosen after initial graphical data analysis as given in the top left of Figure \ref{fig:intro}. In the paper, our candidate set for margins is normal, Student's t with degrees of freedom 3, logistic, log-normal, log-logistic, and gamma distribution. Thus, the algorithm allows also for a heavy-tailed marginal distribution. We can easily incorporate further marginal distributions into our algorithm. However, its current version can estimate the marginal distributions of the real data sets reasonably well, as shown in Section \ref{vcmm-real}. We first truncate a vine tree structure at tree level one for the initial selection of vine copula models, thereby obtaining a Markov tree model. As discussed in Section \ref{vcmm-param-est}, working with Markov trees allows us to decrease the optimization problem's size (CM-step 3) from quadratic to linear in dimension $d$, thereby reducing our computational effort. However, the performance of using different truncation levels might be investigated in the future. A wide range of parametric pair copula families is applicable. Our algorithm's pair copula families are parametric: Gaussian, t, Clayton, Gumbel, Frank, Joe, BB1, BB6, and BB8 copula. Since we apply their possible $90^\circ$, $180^\circ$, $270^\circ$ rotations, the total number of pair copula families utilized is $27$. Chapter 3 of \cite{Czado2019} provides more details about them. For this part, we work with the package {\fontfamily{pcr}\selectfont VineCopula} and mainly its function {\fontfamily{pcr}\selectfont RVineStructureSelect}  \citep{Nagler2019}. The VCMM selects $d \cdot k$ marginal distributions, $k$ Markov tree structures, and $(d -1) \cdot k$ pair copula families at this initial step.

The third step (parameter estimation via the ECM algorithm allowing for Markov (vine) tree structures) updates the VCMM parameters while keeping the marginal distributions, Markov tree structures, and pair copula families fixed until the stopping condition discussed in Section \ref{vcmm-param-est} holds. For each cluster, its mixture weight, pair copula parameters, and marginal parameters are updated per iteration. The total number of updated parameters per iteration is the sum of the total number of pair copula parameters, marginal parameters, and mixture weights over $k$ clusters. Assume the ECM stops after $s$ iterations. The total number of updated parameters up to this point is linear in dimension $d$, the total number of clusters $k$, and iterations $s$. 

Next, we partition the observations into $k$ clusters with the updated posterior probabilities as a temporary clustering assignment. An observation is a member of the cluster, where its posterior probability is highest. 

The marginal distributions and dependence structure of the clusters can change due to the successive ECM steps. Moreover, dependencies can exist in higher tree levels, and accounting for those can improve model power. Thus, we perform a final model selection, and it is based on a full vine specification. The fifth step estimates the model components, including all possible vine tree levels and their parameters, with the fourth step's clustering assignment. The VCMM chooses $d \cdot k$ marginal distributions, $k$ vine tree structures, and $\frac{d \cdot (d -1)}{2} \cdot k$ pair copula families. The total number of updated parameters with this final step is linear in the total number of clusters $k$ and iterations $s$, as in the third step, but is quadratic in dimension $d$ due to the estimation of all possible vine tree levels. Using a full vine specification introduces additional parameter estimates; however, it increases model power as shown in Section \ref{AIS}.

The last step (final clustering assignment based on the full vine specification) assigns the observations to the clusters with the final model's posterior probabilities. 
\begin{algorithm}[H]
\caption{Vine copula mixture models clustering: VCMM}\label{vcmm-clustering-algo}
\begin{algorithmic}[1]
\State \textbf{Input}: $d$-dimensional $n$ observations to cluster $\bm{x}_i = (x_{i,1}, \ldots, x_{i,d})^\top \in \R^d$ for $i=1, \ldots, n$ and total number of clusters $k$.
\State \textbf{Output}: A clustering partition of the observations $\mathcal{C} = \{\mathcal{C}_1, \ldots, \mathcal{C}_k\}$,  estimated model components and parameters of the $j$th cluster defined in Section \ref{vcmm} $\mathcal{\hat{F}}_j$, $\bm{\hat{\gamma}}_j$, $\mathcal{\hat{V}}_j$, $\mathcal{\hat{B}}_j(\mathcal{\hat{V}}_j)$, $\bm{\hat{\theta}}_j\big(\mathcal{\hat{B}}_j(\mathcal{\hat{V}}_j)\big)$, $\hat{\pi}_j$ and final posterior probabilities $r^{(final)}_{i,j}$ for $i=1, \ldots, n$ and $j=1, \ldots, k$.
\For {$j =1, \ldots, k$}
\State \textbf{Step \rom{1}: Initial clustering assignment via a fast clustering algorithm}
\State $\bm{x}^{(0)}_{(j)i_j} \gets  (x^{(0)}_{(j)i_j,1}, \ldots, x^{(0)}_{(j)i_j,d})^\top$ \quad for \quad $i_j = 1, \ldots, n^{(0)}_j$, $\sum\limits_{j=1}^{k} n^{(0)}_j = n$ and $\underset{\forall (j, i_j)}{\bigcup}\, \bm{x}^{(0)}_{(j)i_j} = \underset{\forall i}{\bigcup}\, \bm{x}_i $.
\State \textbf{Step \rom{2}: Initial model selection based on the Step \rom{1} assignment}
\For {$p =1, \ldots, d$}
\Indent
\State $\bm{x}^{(0)}_{p(j)}  \gets  (x^{(0)}_{(j)1,p}, \ldots, x^{(0)}_{(j)n^{(0)}_j,p})^\top$,
\State  $(F^{(0)}_{p(j)}, \bm{\gamma}^{(0)}_{p(j)}) \gets$  $\underset{mar, \bm{\gamma}}{\mathrm{arg\,min\,}}  -2 \cdot \ell^{mar}(\bm{\gamma}; \bm{x}^{(0)}_{p(j)} ) +|\bm{\gamma}|  \cdot \text{log }(n^{(0)}_j), \quad$ where $\ell^{mar}$  is the log-likelihood of the univariate distribution $mar$ for $mar \in$ \{candidate univariate parametric marginal distributions\}, 
\State $\bm{u}^{(0)}_{p(j)} \gets F^{(0)}_{p(j)}(\bm{x}^{(0)}_{p(j)}; \bm{\gamma}^{(0)}_{p(j)})$,
\EndIndent
\EndFor
\State $\bm{u}^{(0)}_{(j)}  \gets  (\bm{u}^{(0)}_{1(j)}, \ldots, \bm{u}^{(0)}_{d(j)})^\top$,
\State $\mathcal{F}^{(0)}_j \gets \{F^{(0)}_{1(j)}, \ldots, F^{(0)}_{d(j)}\}$,
\State bicop $\in$ \{candidate parametric pair copula families\},
\State $\pi^{(0)}_j \gets \frac{n^{(0)}_j}{n} $,
\State $model_{markov} \gets$ RVineStructureSelect ($\bm{u}^{(0)}_{(j)}$ ,  familyset=bicop, trunclevel=1),  where for the component $j$, RVineStructureSelect determines its vine tree structure truncated at tree level 1, associated pair copula families among the set bicop and its parameters as described in Section \ref{vcmm-selection},
\State $\mathcal{V}^{(0)}_j \gets model_{markov}\$Matrix$, $\mathcal{B}^{(0)}_j \gets model_{markov}\$Family$ and $\bm{\theta}^{(0)}_j\gets model_{markov}\$ pars$,
\EndFor
\State \textbf{Step \rom{3}: Parameter estimation via the ECM algorithm allowing for Markov (vine) tree structures}
\While{stop=FALSE}
\State $t \gets 0$,
\For {$j =1, \ldots, k$}
\Indent
\State $ \mathcal{F}^{(t+1)}_j \gets \mathcal{F}^{(0)}_j $, $\mathcal{V}^{(t+1)}_j \gets \mathcal{V}^{(0)}_j $, $\mathcal{B}^{(t+1)}_j \gets \mathcal{B}^{(0)}_j$, 
\State Update $r^{(t+1)}_{i,j}$ as in Equation (\ref{eq:E-step})\quad for \quad $i=1, \ldots, n$,
\State Update $\pi^{(t+1)}_j$,  $\bm{\gamma}^{(t+1)}_j$ and $\bm{\theta}^{(t+1)}_j$ sequentially as in Equations (\ref{eq:CM-step-1}), (\ref{eq:CM-step-2}) and (\ref{eq:CM-step-3}), respectively,
\EndIndent
\EndFor
\State $t \gets t+1$,
\If {the termination criterion in Equation (\ref{eq:stopping}) holds }
\State $s \gets t$, stop=TRUE and \textbf{break}.
\EndIf
\EndWhile
\algstore{myalg}
\end{algorithmic}
\end{algorithm}
\begin{algorithm} [H]                    
\begin{algorithmic}                 
\algrestore{myalg}
\State \textbf{Step \rom{4}: Temporary clustering assignment based on the Markov (vine) tree structure}
\State $\bm{x}_i \in \mathcal{C}_{j^*} \iff j^* = \underset{j=1, \ldots, k}{\mathrm{arg\,max\,}} r^{(s)}_{i,j}$ \quad for \quad $i =1, \ldots, n$.
\For {$j =1, \ldots, k$}
\State $\bm{x}^{(s)}_{(j)i_j} \gets  (x^{(s)}_{(j)i_j,1}, \ldots, x^{(s)}_{(j)i_j,d})^\top$ \quad for \quad $i_j = 1, \ldots, n^{(s)}_j$, $\sum\limits_{j=1}^{k} n^{(s)}_j = n$ and $\underset{\forall (j, i_j)}{\bigcup}\, \bm{x}^{(s)}_{(j)i_j} = \underset{\forall i}{\bigcup}\, \bm{x}_i $.
\State \textbf{Step \rom{5}: Final model selection based on a full vine specification}
\State Perform lines 7-- 12 with the new assignment $\bm{x}^{(s)}_{(j)i_j}$ and change the iteration index from $(0)$ to $(s)$,
\State $model_{VCMM} \gets$ RVineStructureSelect ($\bm{u}^{(s)}_{(j)}$, familyset=bicop, trunclevel=d-1),
\State $\mathcal{V}^{(s)}_j \gets model_{VCMM}\$Matrix$, $\mathcal{B}^{(s)}_j \gets model_{VCMM}\$Family$ and $\bm{\theta}^{(s)}_j\gets model_{VCMM}\$pars$.
\EndFor
\State \textbf{Step \rom{6}: Final clustering assignment based on the full vine specification}
\State $\bm{x}_i \in \mathcal{C}_{j^*} \iff j^* = \underset{j=1, \ldots, k}{\mathrm{arg\,max\,}} r^{(s+1)}_{i,j}$ for $i =1, \ldots, n$, where $r^{(s+1)}_{i,j}$ are the posterior probabilities calculated from the final model, 
\State $\mathcal{\hat{F}}_j \gets \{F^{(s)}_{1(j)}, \ldots, F^{(s)}_{d(j)}\}$,  $\hat{\bm{\gamma}}_j \gets \{\bm{\gamma}^{(s)}_{1(j)}, \ldots,  \bm{\gamma}^{(s)}_{d(j)}\}$, $\mathcal{\hat{V}}_j \gets \mathcal{V}^{(s)}_j$, $\mathcal{\hat{B}}_j(\mathcal{\hat{V}}_j) \gets \mathcal{B}^{(s)}_j$, $\bm{\hat{\theta}}_j\big(\mathcal{\hat{B}}_j(\mathcal{\hat{V}}_j)\big) \gets \bm{\theta}^{(s)}_j$, $\hat{\pi}_j \gets \pi^{(s)}_j$, $r^{(final)}_{i,j} \gets r^{(s+1)}_{i,j}$.
\end{algorithmic}
\end{algorithm}
\section{Simulation studies}\label{vcmm-sim}
This section will demonstrate the remarkable and promising results of our clustering algorithm, VCMM, using simulated data. We compare its performance with the initial partition from k-means to some well-known model-based clustering algorithms: the mixture of multivariate normal, skew normal, t, and skew t distributions. We fit the mixture of multivariate normal distributions using the package {\fontfamily{pcr}\selectfont mclust} and the others using the package {\fontfamily{pcr}\selectfont mixsmsn}. The latter fits scale mixtures of skew normal distributions and works with the initial partition of k-means like our algorithm. Therefore, the comparison of the performance of the VCMM and the chosen competing algorithms is fairer. We work with the default specifications of the packages but specify the total number of clusters and seed. For the clustering performance evaluation, we use the BIC value and misclassification rate when the true labels are available. The lower the BIC value or misclassification rate, the better the clustering assignment. 

The BIC criterion compares only models studied; thus, it selects a better model among the evaluated ones. However, it does not imply that it selects the best model. Since unsupervised learning problems do not contain the true labels of the observations, a separation of the data in training and test sets is not feasible. Therefore, the misclassification rate that we consider can be regarded as the in-sample misclassification rate of supervised learning. 

In Sections \ref{gauss-nongauss} and \ref{nongauss-nongauss}, the simulated data has three variables, two clusters, either $100$ or $500$ observations in each cluster, and we replicate their data generating process $100$ times. $X_{p(1)} = F^{-1}_{p(1)}(U_{p(1)}; \bm{\gamma}_{p(1)})$ and $X_{p(2)} = F^{-1}_{p(2)}(U_{p(2)}; \bm{\gamma}_{p(2)})$ for $p =1,2,3$ give the variables of the first and second clusters, respectively. We simulate $U_{p(1)}$ and $U_{p(2)}$ from a specified vine copula model and specify $F_{p(1)}, F_{p(2)}, \bm{\gamma}_{p(1)}, \bm{\gamma}_{p(2)}$. A mixture of vine copulas generates the first scenario in Section \ref{gauss-nongauss} with one/two parameter pair copula families and non-Gaussian margins. Section \ref{nongauss-nongauss} simulates data from a mixture of vine copulas with single parameter pair copulas and Gaussian/non-Gaussian margins. In the first scenario, we aim to analyze how well the VCMM improves the clustering compared to its starting partition from k-means. In the second scenario, we aim to analyze how well the VCMM and other model-based clustering algorithms capture different dependence structures and shapes hidden in the multivariate data with different numbers of observations. 

In Section \ref{vcmm-sim-4}, we discuss the effect of different initial clustering techniques at Step \rom{1} in Algorithm \ref{vcmm-clustering-algo} on the VCMM using the artificial data from a mixture of vine copulas with Gaussian copulas and Gaussian margins. In Section \ref{misspecification}, we aim to analyze the performance of the VCMM when the data generating process is misspecified, e.g., we simulate the data from a mixture of multivariate skew t distributions. 
\subsection{The mixture of vine copulas with one/two parameter pair copulas and non-Gaussian margins}\label{gauss-nongauss}
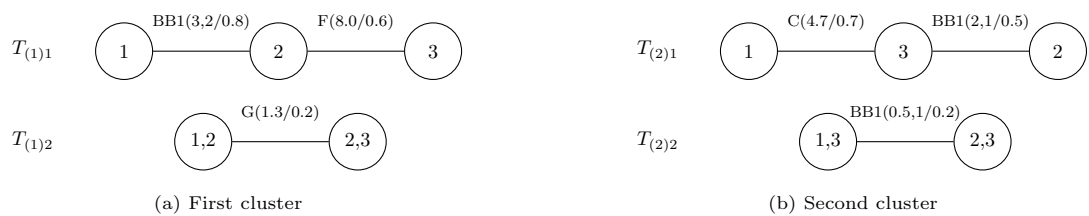
\begin{figure}[ht]
\begin{subfigure}[b]{0.45\linewidth}
	\centering
	\renewcommand{\xshiftNodes}{0.15*\linewidth}
	\renewcommand{\yshiftLabels}{.0cm}  
\begin{tikzpicture}	[every node/.style = VineNode, node distance =1.6cm,scale=0.75, transform shape]
\node[TreeLabels] (T1) {$T_{(1)1}$}
node             (v1)         [right of = T1] {1}			
node             (v2)         [right of = v1, xshift = \xshiftNodes] {2}
node             (v3)         [right of = v2, xshift = \xshiftNodes] {3};
\draw[color=black] (v1) to node[draw=none, font = \labelsize, fill = none, above, yshift = \yshiftLabels] {BB1(3,2/0.8)} (v2);
\draw[color=black] (v2) to node[draw=none,  font = \labelsize,fill = none, above, yshift = \yshiftLabels] {F(8.0/0.6)} (v3);
\node[TreeLabels] (T2)      [below of = T1] {$T_{(1)2}$}
node             (v13)         [right of = T2, xshift = 1.25*\xshiftNodes] {1,2}
node             (v23)         [right of = v13, xshift =\xshiftNodes] {2,3};
\draw[color=black] (v13) to node[draw=none, font = \labelsize, fill = none, above, yshift = \yshiftLabels] {G(1.3/0.2)} (v23);
\end{tikzpicture}
\caption{First cluster}
\end{subfigure}
\qquad
\begin{subfigure}[b]{0.45\linewidth}
	\centering
	\renewcommand{\xshiftNodes}{0.15*\linewidth}
	\renewcommand{\yshiftLabels}{.0cm}  
\begin{tikzpicture}	[every node/.style = VineNode, node distance =1.6cm,scale=0.75, transform shape]
\node[TreeLabels] (T1)    {$T_{(2)1}$}
node             (v1)         [right of = T1] {1}			
node             (v2)         [right of = v1, xshift = \xshiftNodes] {3}
node             (v3)         [right of = v2, xshift =\xshiftNodes] {2};
\draw[color=black] (v1) to node[draw=none,  font = \labelsize,fill = none, above, yshift = \yshiftLabels] {C(4.7/0.7)} (v2);
\draw[color=black] (v2) to node[draw=none, font = \labelsize, fill = none, above, yshift = \yshiftLabels] {BB1(2,1/0.5)} (v3);
\node[TreeLabels] (T2)      [below of = T1] {$T_{(2)2}$}
node             (v13)         [right of = T2, xshift = 1.25*\xshiftNodes] {1,3}
node             (v23)         [right of = v13, xshift =\xshiftNodes] {2,3};
\draw[color=black] (v13) to node[draw=none,  font = \labelsize,fill = none, above, yshift = \yshiftLabels] {BB1(0.5,1/0.2)} (v23);
\end{tikzpicture}
\caption{Second cluster}
\end{subfigure}
\caption{Vine tree structure of simulated data with one/two parameter pair copulas and two clusters. A letter at an edge refers to its bivariate copula family, where C: Clayton, G: Gumbel,  F: Frank, and BB1: BB1 copula. The true parameter value(s) and corresponding Kendall's $\tau$ of the pair copula are given inside the parenthesis (parameter(s)/Kendall's $\tau$).}
\label{fig:vcmm-case-1}
\end{figure}

We simulate data $U_{p(1)}$ and $U_{p(2)}$ for $p =1,2,3$ from a vine copula model, where pair copula families have either single or two parameters, as shown in Figure \ref{fig:vcmm-case-1}. Both clusters include positive as well as high, medium, and low strength dependencies.

\begin{figure}[H]
\begin{subfigure}[b]{0.45\linewidth}
\centering
\includegraphics[width=.7\textwidth]{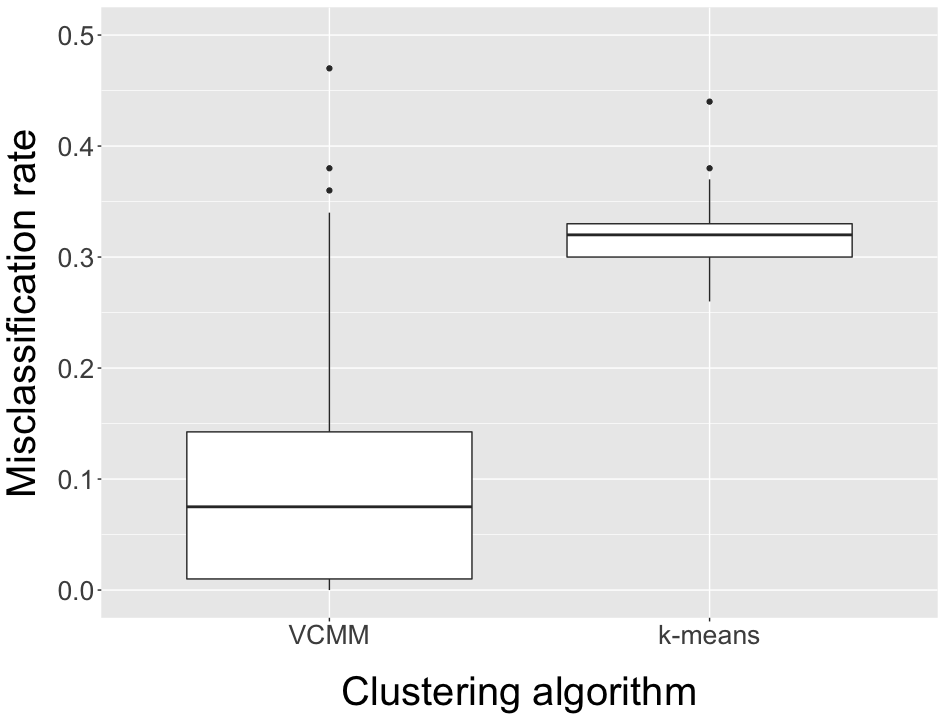} 
\caption{$500$ observations per cluster}
\end{subfigure}
\qquad
\begin{subfigure}[b]{0.45\linewidth}
\centering
\includegraphics[width=.7\textwidth]{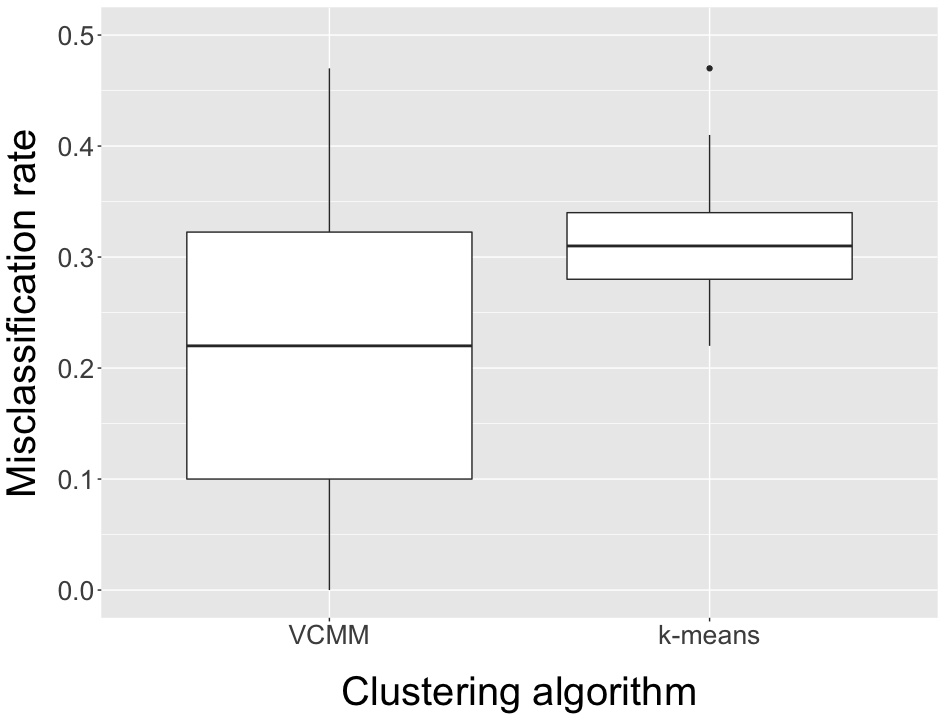}
\caption{$100$ observations per cluster}
\end{subfigure}
\caption{Comparison of the clustering performance of the VCMM and its initial partition algorithm k-means over $100$ replications under the scenario specified in Figure \ref{fig:vcmm-case-1} and Table \ref{table:vcmm-case-1}.}
\label{fig:vcmm-case-1-box}
\end{figure}

Then we transform the data from u-scale $U_{p(1)}, U_{p(2)}$ to x-scale $X_{p(1)}, X_{p(2)}$ as explained in Section \ref{vcmm-sim}. Table \ref{table:vcmm-case-1} presents the marginal distributions with the parameters. The clusters are non-Gaussian.
\begin{table}[H]
\centering
\scalebox{0.75}{
\begin{tabular}{l  l l  l l  l}
\hline
 $F_{1(1)}(\bm{\gamma}_{1(1)})$ &  $F_{2(1)}(\gamma_{2(1)})$&  $F_{3(1)}(\gamma_{3(1)})$&  $F_{1(2)}(\bm{\gamma}_{1(2)})$&  $F_{2(2)}(\bm{\gamma}_{2(2)})$&  $F_{3(2)}(\gamma_{3(2)})$ \\ 
 $llogis(1.5,1.25)$& $exp(0.1)$& $lnorm(0.1,1.3)$& $lnorm(2.5,0.5)$  &   $logis(5,3)$&  $exp(0.05)$\\ 
  \hline
\end{tabular}}
\caption{Univariate marginal distributions and associated parameters of each cluster. They are defined in \ref{app-margin}.} 
\label{table:vcmm-case-1}
\end{table}
We evaluate the clustering performance of the VCMM and k-means, visualizing the misclassification rate per simulation replication in box plots. For the larger number of observations  ($500$ observations per cluster), Figure \ref{fig:vcmm-case-1-box} shows that the VCMM provides a noticeably better fit than k-means. On average, it improves the accuracy by $22\%$ compared to its initial partition. For the small number of observations ($100$ observations per cluster), its average misclassification rate is $10\%$ less than k-means, but the VCMM variance in the accuracy increases as the number of observations gets lower. The VCMM requires, on average, 25 and 17 ECM iterations for the large and small numbers of observations, respectively, i.e., the convergence takes more iterations with a larger data set here.
\subsection{The mixture of vine copulas with one parameter pair copulas and Gaussian/non-Gaussian margins}\label{nongauss-nongauss}
\begin{figure}[H]
\begin{subfigure}[b]{0.45\linewidth}
	\centering
	\renewcommand{\xshiftNodes}{0.15*\linewidth}
	\renewcommand{\yshiftLabels}{.0cm}  
\begin{tikzpicture}	[every node/.style = VineNode, node distance =1.6cm,scale=0.75, transform shape]
\node[TreeLabels] (T1) {$T_{(1)1}$}
node             (v1)         [right of = T1] {1}			
node             (v2)         [right of = v1, xshift = \xshiftNodes] {2}
node             (v3)         [right of = v2, xshift = \xshiftNodes] {3};
\draw[color=black] (v1) to node[draw=none,  font = \labelsize,fill = none, above, yshift = \yshiftLabels] {G(2.5/0.6)} (v2);
\draw[color=black] (v2) to node[draw=none,  font = \labelsize,fill = none, above, yshift = \yshiftLabels] {SG(5.0/0.8)} (v3);
\node[TreeLabels] (T2)      [below of = T1] {$T_{(1)2}$}
node             (v13)         [right of = T2, xshift = 1.25*\xshiftNodes] {1,2}
node             (v23)         [right of = v13, xshift =\xshiftNodes] {2,3};
\draw[color=black] (v13) to node[draw=none,  font = \labelsize,fill = none, above, yshift = \yshiftLabels] {C(0.9/0.3)} (v23);
\end{tikzpicture}
\caption{First cluster}
\end{subfigure}
\qquad
\begin{subfigure}[b]{0.45\linewidth}
	\centering
	\renewcommand{\xshiftNodes}{0.15*\linewidth}
	\renewcommand{\yshiftLabels}{.0cm}  
\begin{tikzpicture}	[every node/.style = VineNode, node distance =1.6cm,scale=0.75, transform shape]
\node[TreeLabels] (T1)   {$T_{(2)1}$}
node             (v1)         [right of = T1] {1}			
node             (v2)         [right of = v1, xshift = \xshiftNodes] {2}
node             (v3)         [right of = v2, xshift = \xshiftNodes] {3};
\draw[color=black] (v1) to node[draw=none, font = \labelsize, fill = none, above, yshift = \yshiftLabels] {F(11.4/0.7)} (v2);
\draw[color=black] (v2) to node[draw=none,  font = \labelsize,fill = none, above, yshift = \yshiftLabels] {SC(2.0/0.5)} (v3);
\node[TreeLabels] (T2)      [below of = T1] {$T_{(2)2}$}
node             (v13)         [right of = T2, xshift =1.25*\xshiftNodes] {1,2}
node             (v23)         [right of = v13, xshift =\xshiftNodes] {2,3};
\draw[color=black] (v13) to node[draw=none,  font = \labelsize,fill = none, above, yshift = \yshiftLabels] {J(1.4/0.2)} (v23);
\end{tikzpicture}
\caption{Second cluster}
\end{subfigure}
\caption{Vine tree structure of simulated data with two clusters. A letter at an edge refers to its bivariate copula family, where C: Clayton, SC: Survival Clayton, G: Gumbel,  SG: Survival Gumbel, F: Frank, and J: Joe copula. The true parameter value and corresponding Kendall's $\tau$ of the pair copula are given inside the parenthesis (parameter/Kendall's $\tau$).}
\label{fig:vcmm-case-2}
\end{figure}
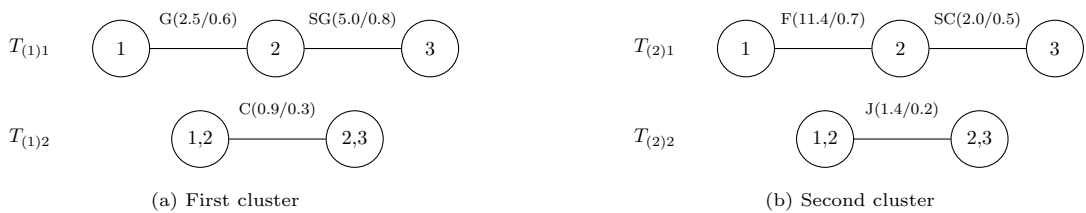

\begin{figure}[H]
\centering
\includegraphics[width=.3\textwidth]{intro-x.png}  \hspace{2cm}
\includegraphics[width=.3\textwidth]{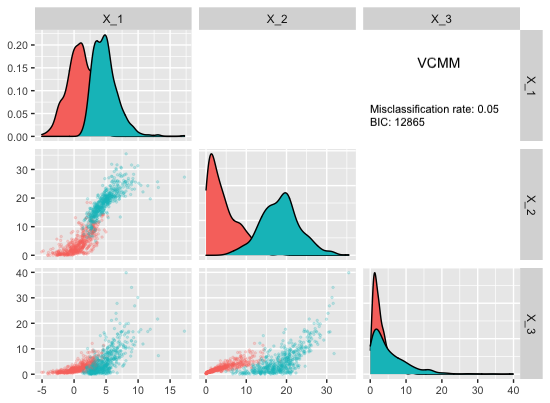}
\caption{Pairwise scatter plot of simulated data ($500$ observations per cluster) on x-scale under the scenario specified in Figure \ref{fig:vcmm-case-1} and Table \ref{table:vcmm-case-1} (left). The right plot shows the fitted VCMM. The red and green points show the observations of one cluster and the other cluster, respectively. The diagonal of the plots shows each cluster's associated variable's marginal density function.}
\label{fig:vcmm-case-2-x}
\end{figure}

\begin{figure}[H]
\centering
\includegraphics[width=.3\textwidth]{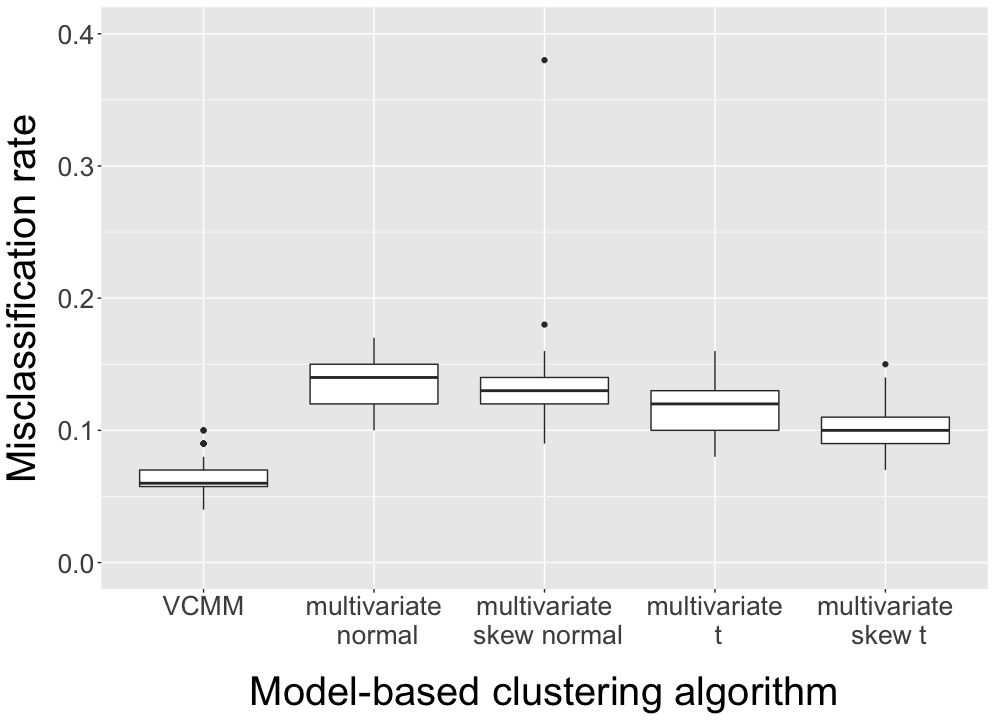}  \hspace{2cm}
\includegraphics[width=.3\textwidth]{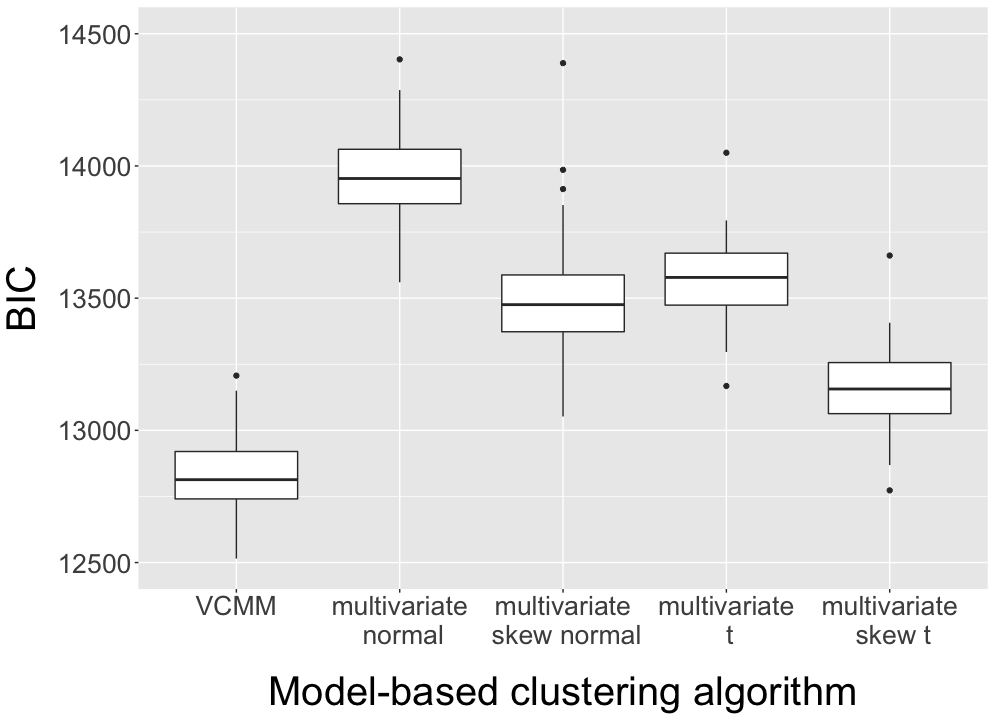}
\caption{Comparison of the model-based clustering algorithms' performance over $100$ replications with $500$ observations per cluster under the scenario specified in Figure \ref{fig:vcmm-case-2} and Table \ref{table:vcmm-case-2}.}
\label{fig:vcmm-case-2-box}
\end{figure}

\begin{figure}[ht]
\centering
\includegraphics[width=.3\textwidth]{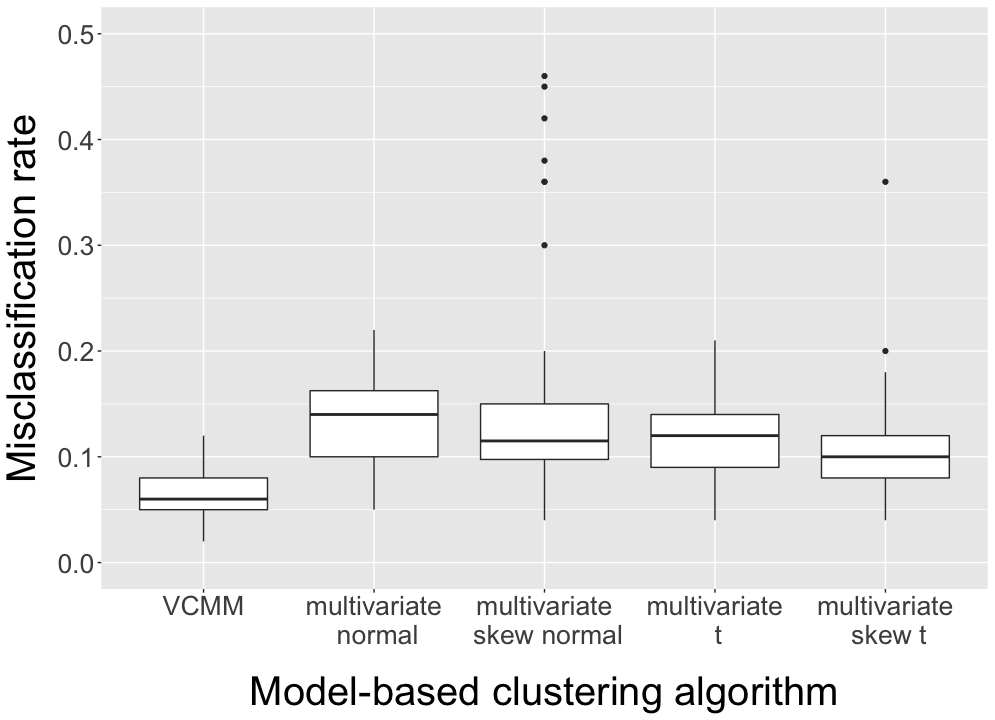}  \hspace{2cm}
\includegraphics[width=.3\textwidth]{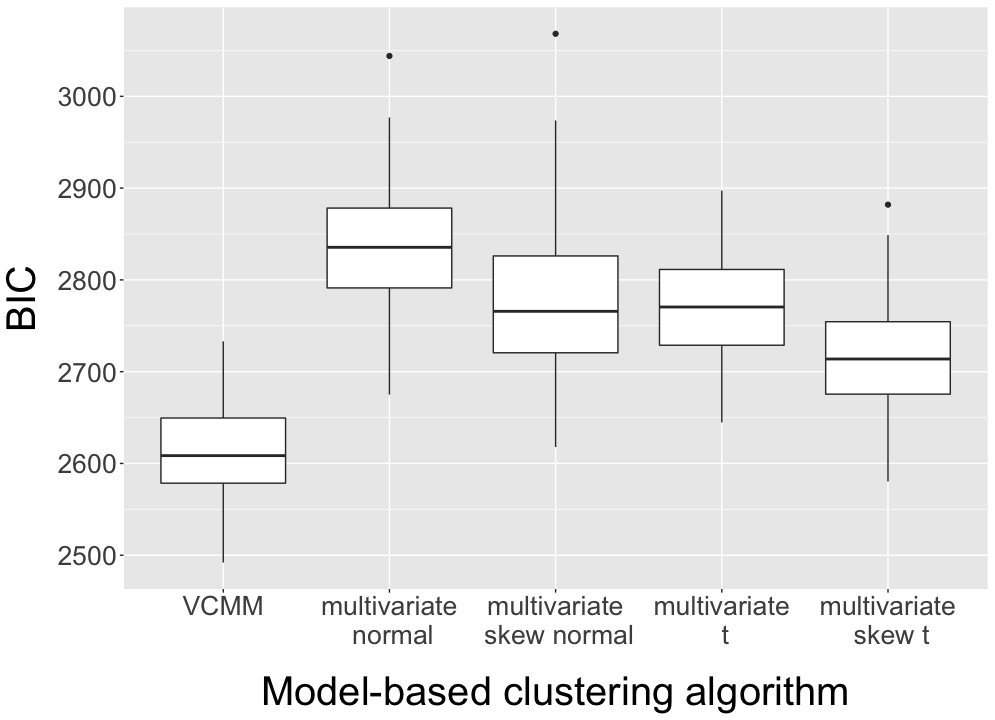}
\caption{Comparison of the model-based clustering algorithms' performance over $100$ replications with $100$ observations per cluster under the scenario specified in Figure \ref{fig:vcmm-case-2} and Table \ref{table:vcmm-case-2}.}
\label{fig:vcmm-case-2-box-n100}
\end{figure}

In this scenario, we work with single parameter pair copulas, add Gaussian margins within the clusters, simulate data on u-scale $U_{p(1)}$, $U_{p(2)}$ for $p =1,2,3$ from the vine tree structures shown in Figure \ref{fig:vcmm-case-2}. Both clusters share the same vine tree structure and include asymmetric tail dependencies. The second cluster has a symmetric, non-Gaussian dependency (Frank copula) between the variables $U_{1(2)}$ and $U_{2(2)}$.
The next step is to obtain the data on the x-scale $X_{p(1)}, X_{p(2)}$ as described before. Marginal distributions and the parameters used are listed in Table \ref{table:vcmm-case-2}.  We illustrate simulated data ($500$ observations per cluster) on the x-scale in the left panel of Figure \ref{fig:vcmm-case-2-x}, where a color represents the observations from a cluster. The generated clusters are non-elliptical. The fitted VCMM detects the true shape of the clusters as shown in the right panel of Figure \ref{fig:vcmm-case-2-x} as opposed to other model-based clustering algorithms given in Figure \ref{fig:intro}.
\begin{table}[H]
\centering
\scalebox{0.75}{
\begin{tabular}{l  l l  l l  l}
\hline
 $F_{1(1)}(\bm{\gamma}_{1(1)})$ &  $F_{2(1)}(\gamma_{2(1)})$&  $F_{3(1)}(\gamma_{3(1)})$&  $F_{1(2)}(\bm{\gamma}_{1(2)})$&  $F_{2(2)}(\bm{\gamma}_{2(2)})$&  $F_{3(2)}(\gamma_{3(2)})$ \\ 
 $\mathcal{N}(1,2)$ & $exp(0.2)$& $lnorm(0.8, 0.8)$& $lnorm(1.5,0.4)$ &  $\mathcal{N}(18,5)$&  $exp(0.2)$\\ 
  \hline
\end{tabular}}
\caption{Univariate marginal distributions and associated parameters of each cluster. They are defined in \ref{app-margin}.} 
\label{table:vcmm-case-2}
\end{table}
Figure \ref{fig:vcmm-case-2-box} visualizes the misclassification rate and BIC value per simulation replication in box plots for 500 observations per cluster. The VCMM is superior to other model-based clustering algorithms regarding both the misclassification rate and the BIC value. It separates two non-elliptical clusters adequately as expected, whereas others tend to find it challenging. Even though they cannot model two clusters as the VCMM does, the mixtures of multivariate skew distributions show a better fit than elliptical distributions in terms of the BIC value. The mean misclassification rate of the mixtures of multivariate skew t distributions and that of the VCMM are $4\%$ and $8\%$ less than that of multivariate normal distributions.
Figure \ref{fig:vcmm-case-2-box-n100} shows that the misclassification rates of the VCMM do not change enormously for $100$ observations per cluster and are still lower than the others. Its variance in the accuracy increases with the smaller number of observations as in the first simulation scenario. The BIC values favor the VCMM, and after the VCMM, the mixtures of multivariate skew distributions provide better fits than the others. 
\subsection{The mixture of vine copulas with Gaussian copulas and Gaussian margins: significant overlaps}\label{vcmm-sim-4}
\begin{figure}[ht]
\centering
\includegraphics[width=.3\textwidth]{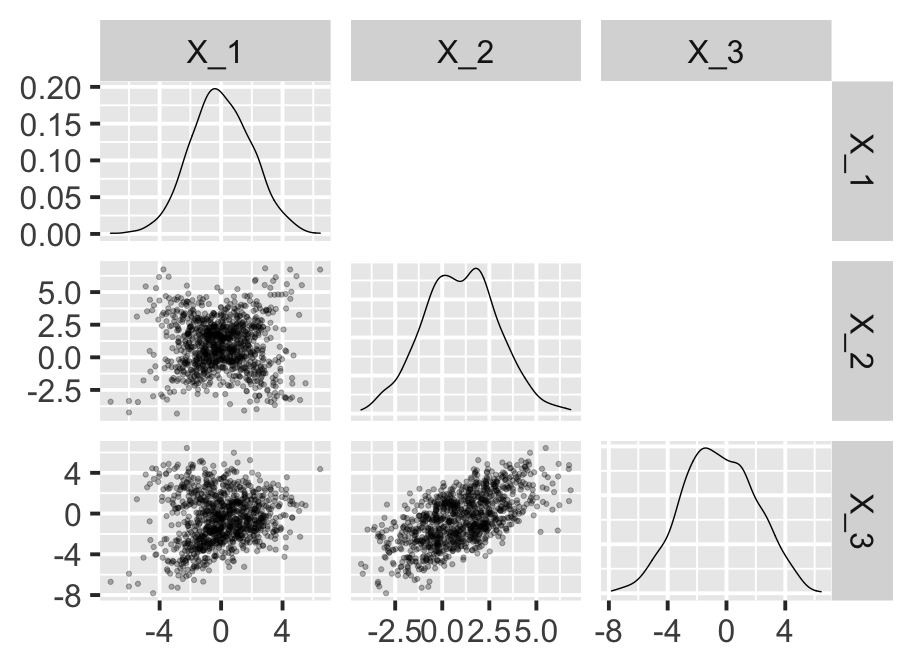} \hspace{2cm}
\includegraphics[width=.3\textwidth]{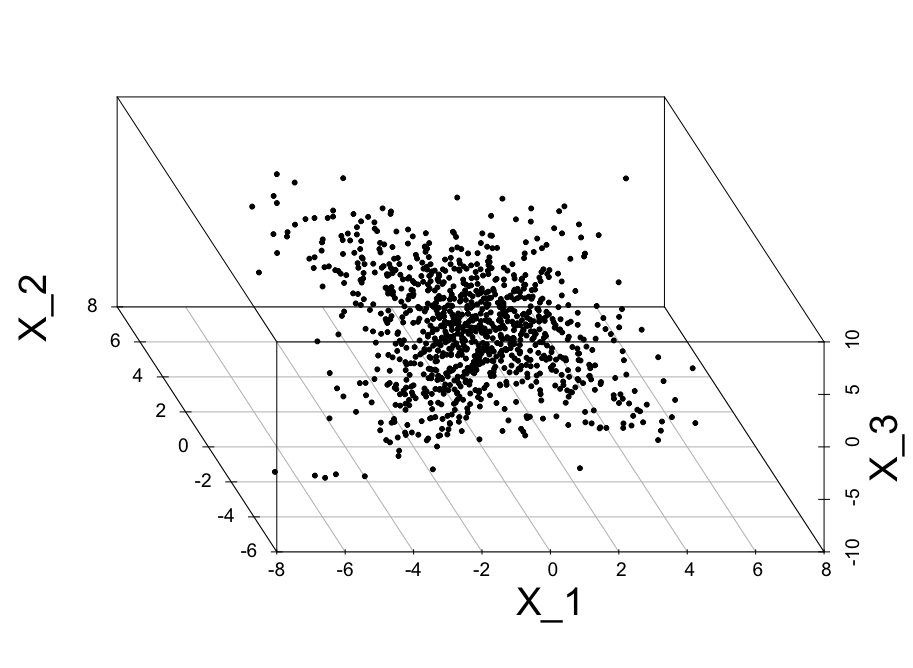}
\caption{Pairwise(left) and $3$-dimensional scatter plot(right) of artificial data ($500$ observations per cluster) on x-scale under the scenario specified in Figure \ref{fig:vcmm-case-3} and Table \ref{table:vcmm-case-3}. The diagonal of the plot on the left shows the corresponding variable's marginal density function.}
\label{fig:vcmm-case-4-x}
\end{figure}

We generate artificial data from a mixture of two vine copulas as specified in \ref{app-sim}. They have only Gaussian copulas between the pairs of variables, and each margin follows a univariate normal distribution. Thus, the scenario represents a mixture of multivariate normal distributions. We do not observe strong multimodality of variables in the diagonal of the left panel in Figure \ref{fig:vcmm-case-4-x}. However, the data has two obvious clusters with many overlaps, creating a x shape in three dimensions, in the right panel of Figure \ref{fig:vcmm-case-4-x}. 

In an attempt to determine the true groups, we fit the VCMM as in previous simulation studies. Since k-means assumes spherical shapes of clusters, we suspect that the VCMM's model selection using the initial partition from k-means could be satisfactory in such a scenario. The VCMM results in $65\%$ classification accuracy with the BIC of 12447.  As we suggest at Step \rom{1} in Algorithm \ref{vcmm-clustering-algo}, other partition strategies can be used before fitting the VCMM. For instance, we partition the data by the model-based hierarchical clustering using the function {\fontfamily{pcr}\selectfont hcVVV} with its default specifications in the package {\fontfamily{pcr}\selectfont mclust} and then run our algorithm. As a result, the VCMM has $95\%$ classification accuracy with the BIC of 12039. Therefore, using different starting partitions in the VCMM and selecting a final model with the lowest BIC are suggested. Also, fitting the mixture of multivariate normal distributions gives $96\%$ classification accuracy with the BIC of 11987. 

To conclude, the VCMM does show its effectiveness and flexibility in clustering multivariate non-Gaussian and Gaussian data in our simulation studies when the data generating process is a mixture of vine copulas. Even though its starting partition cannot reasonably identify the clusters, the VCMM often deals with them.
\subsection{The mixture of multivariate skew t distributions}\label{misspecification}
As a misspecification scenario, we simulate data in three dimensions with two components from the mixture of multivariate skew t distributions expressed as a class of skew normal independent distributions \citep{Cabral2012}. The total number of observations is 1000. Denoting the density of a multivariate skew t distribution by $ST(.; \bm{\mu}, \bm{\Sigma}, \bm{\lambda}, \nu)$ with location vector $\bm{\mu}$, scale matrix $\bm{\Sigma}$, skewness vector $\bm{\lambda}$, and degrees of freedom $\nu$, assume a random vector $\bm{X}$ has this mixture distribution. Then its density at $\bm{x}$ is given by 
\begin{equation}
\pi_1 \cdot ST(\bm{x}; \bm{\mu}_1, \bm{\Sigma}_1, \bm{\lambda}_1, \nu_1) + (1 - \pi_1) \cdot ST(\bm{x}; \bm{\mu}_1, \bm{\Sigma}_2, \bm{\lambda}_2, \nu_2),
\label{eq:miss}
\end{equation}
where the true values of the parameters used in the simulation are $\bm{\mu}_1 =(1,1,0)^\top$, $\bm{\mu}_2 =(-2,-2,-2)^\top$, $\pi_1=0.6$, $\bm{\lambda}_1=(4,-4,4)^\top$, $\bm{\lambda}_2=(-4,4,4)^\top$, $ \nu_1=8$, $ \nu_2=10$, and $\bm{\alpha}_1 = \bm{\alpha}_2 = (2,1,1,2,1,2)$ vectorizing the upper triangular matrix of symmetric matrices $\bm{\Sigma}_1 = \bm{\Sigma}_2$. We replicate the data generating process 100 times, and Table \ref{table:misspecification-result} shows that the mixture of multivariate skew t distributions fits well with the data as expected. The mixture of multivariate normal distributions has more difficulty in capturing non-elliptical components than the VCMM. All in all, the VCMM has good credibility also under this scenario.
\begin{table}[H]
\centering
\scalebox{0.75}{
\begin{tabular}{l  c  c  c  l}
Mixture model & Multivariate skew t & VCMM& Multivariate normal \\
\hline
Average misclassification rate & \textbf{0.06}&0.09&0.13\\
Average BIC&\textbf{10676}&10822&10881 \\
Average log-likelihood &\textbf{-5248}&-5330&-5380\\
  \hline
\end{tabular}}
\caption{Comparison of the model-based clustering algorithms' performance over $100$ replications with 1000 observations under the scenario specified in Equation \eqref{eq:miss}. The best result in each row is highlighted.} 
\label{table:misspecification-result}
\end{table}

\section{Real data sets}\label{vcmm-real}
This section illustrates the VCMM's usefulness in analyzing and clustering multivariate non-Gaussian real data. We first focus on improvements the VCMM provides thanks to its steps given in Algorithm \ref{vcmm-clustering-algo}. Later, we analyze the effect of using a fixed vine tree structure for clusters in the VCMM and compare its performance with other models. We run other clustering algorithms using the same random seed for a fair comparison, thereby using the same starting partitions (except the mixture of multivariate normal distributions). We report their results regarding the misclassification rate and BIC value as in the simulation studies in Section \ref{vcmm-sim}. 

Additionally, we will discuss the computational effort of the VCMM. All computations are run on a MacBook Air ($2018$) with a $1,6$ GHz Dual-Core Intel Core i5 and 8 GB of RAM, running {\fontfamily{pcr}\selectfont R} version $4.0.3$. Section \ref{unsupervised} will also study the issue of determining the optimal number of clusters in the VCMM.
\subsection{AIS}\label{AIS}
\begin{figure}[H]
        \centering
        \begin{subfigure}[b]{0.4\textwidth}
            \centering
            \includegraphics[width=\textwidth]{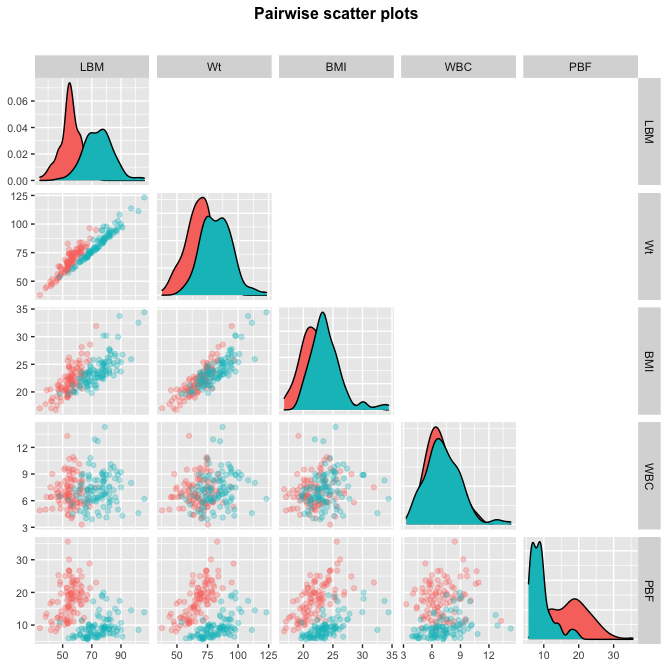}
        \end{subfigure}
        \hspace{2cm}
        \begin{subfigure}[b]{0.4\textwidth}  
            \centering 
            \includegraphics[width=\textwidth]{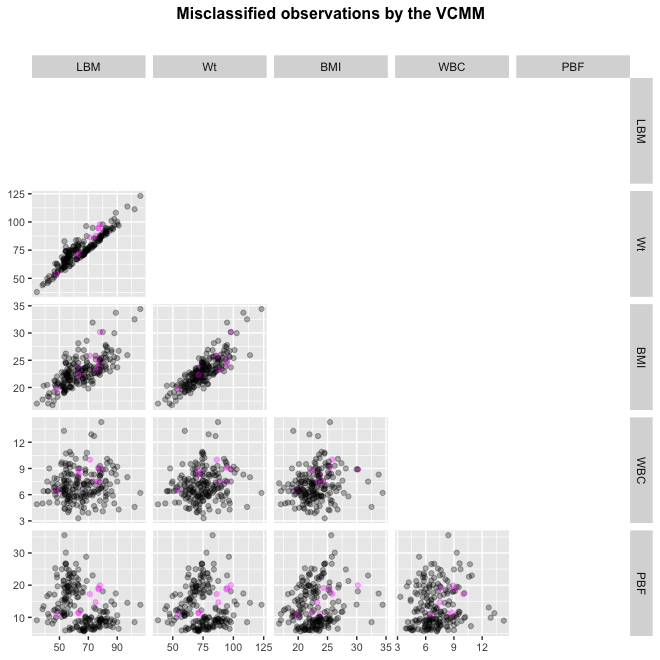}
        \end{subfigure}
        \vskip\baselineskip
        \begin{subfigure}[b]{0.4\textwidth}   
            \centering 
            \includegraphics[width=\textwidth]{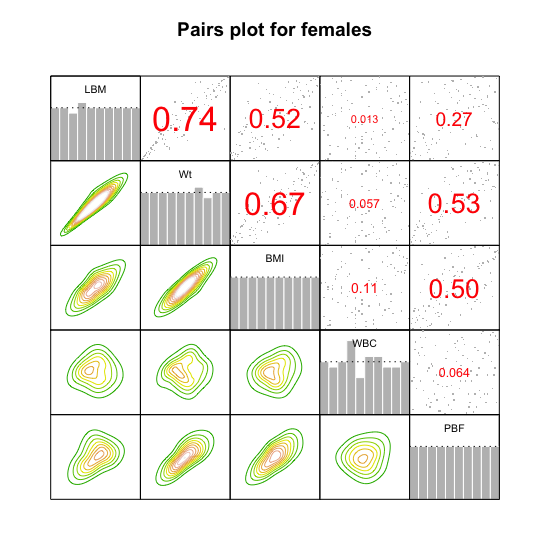}
        \end{subfigure}
        \hspace{2cm}
        \begin{subfigure}[b]{0.4\textwidth}   
            \centering 
            \includegraphics[width=\textwidth]{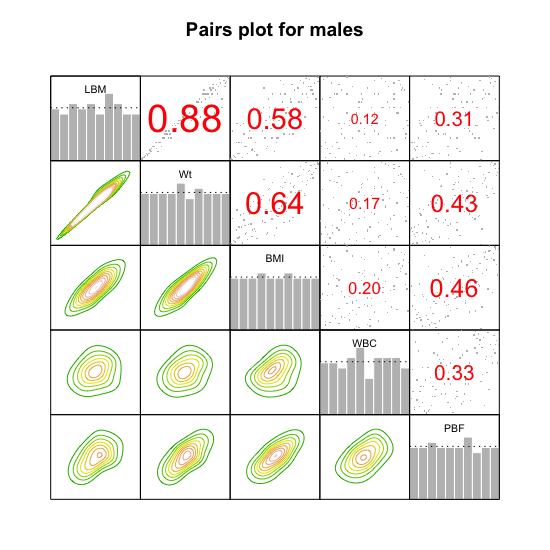}
        \end{subfigure}
 \caption{Pairwise scatter plot of the subset of {\fontfamily{pcr}\selectfont AIS} data (top left) with red points for observations of females and green points for males, where diagonal: marginal density function of each class's corresponding variable, and of the misclassified observations by the VCMM shown by magenta (top right). Pairs plots of females (bottom left) and males (bottom right), where upper: pairs plots of copula data, diagonal: histogram of copula margins, lower: normalized contour plots.}
\label{fig:AIS-x}
\end{figure}

A well-analyzed Australian Institute of Sport (AIS) data consists of $13$ measurements made on $102$ male and $100$ female athletes. Our objective of clustering this data is to see if the VCMM can find two clusters for females and males and analyze its steps' performance. For our analysis, we select a subset of five variables: lean body mass (LBM), weight (Wt), body mass index (BMI), white blood cell count (WBC), and percentage of body fat (PBF). This sample appears non-Gaussian and has asymmetric dependence patterns shown on the bottom panels in Figure \ref{fig:AIS-x}. To obtain u-data, we fit the empirical cumulative distribution function of each variable in each class. The lower panels in pairs plots show normalized contour plots, and we often do not observe Gaussian dependence since most contours are non-elliptical. The pairwise dependence between the same pair of variables usually is the same in females and males, but its strength is different (e.g., Wt and BMI). The marginal density function of BMI and WBC for both classes is similar to each other as shown in the diagonal of the top left panel in Figure \ref{fig:AIS-x}.

Fitting k-means with 10000 different seeds leads to two different partitions of the data set. We fit the VCMM using both partitions but present the result for the best VCMM, whose BIC value is lower than the other. However, both models have the same final accuracy.

Since we know the gender of each observation, we can evaluate the misclassification rate of the binary classification for the clustering algorithm and associate the final clusters with the classes. The VCMM improves its clustering power with its steps in Algorithm \ref{vcmm-clustering-algo} as shown in Table \ref{table:AIS-result-vcmm}. The starting partition obtained from k-means assigns almost one-fourth of males to females (Step \rom{1}). Then Step \rom{2} fits the initial VCMM model using Markov trees, thereby returning a log-likelihood value. The accuracy is eight percent higher than the starting partition. After the ECM iterations and temporary clustering assignment (Step \rom{3} and Step \rom{4}), the VCMM reduces the misclassification rate by $12\%$ compared to k-means. Selecting the final VCMM with a full vine specification (Step \rom{5} and Step \rom{6}) provides a crucial gain in the log-likelihood. In the end, the VCMM identifies almost all females correct except one female. Overall it provides $12\%$ higher accuracy in revealing females and males than k-means. The misclassified observations by the VCMM lie on the boundary of the classes as shown in the top right panel of Figure \ref{fig:AIS-x}.
\begin{table}[H]
\centering
\scalebox{0.75}{
\begin{tabular}{l p{.5cm} p{.5cm} p{.5cm}  p{.5cm}  p{.5cm} p{.5cm} p{.5cm} p{.5cm}}
Clustering algorithm & \multicolumn{2}{c}{k-means} & \multicolumn{2}{c}{\begin{tabular}[c]{@{}c@{}}VCMM after Step II \\ (with Markov trees)\end{tabular}} & \multicolumn{2}{c}{\begin{tabular}[c]{@{}c@{}}VCMM after Step IV \\ (ECM using Markov trees)\end{tabular}} & \multicolumn{2}{c}{\begin{tabular}[c]{@{}c@{}}final VCMM \\ (Full vine using Step IV's assignment)\end{tabular}} \\ \hline
Gender & 1 & 2 & \,\,\,\,\,\,\,\,\,\,\,\,\,\,\,1 & \,\,\,\,\,\,\,\,\,\,\,\,\,\,\,2 & \,\,\,\,\,\,\,\,\,\,\,\,\,\,\,\,\,\,\,\,\,\,\,\,1 & \,\,\,\,\,\,\,\,\,\,\,\,\,\,\,\,\,\,\,\,\,\,\,\,2 & \,\,\,\,\,\,\,\,\,\,\,\,\,\,\,\,\,\,\,\,\,\,\,\,\,\,\,\,\,\,\,\,\,\,\,\,\,\,1 & \,\,\,\,\,\,\,\,\,\,\,\,\,\,\,\,\,\,\,\,\,\,\,\,\,\,\,\,\,\,\,\,\,\,\,\,\,\,2 \\ \hline
Female & 93 & 7 & \,\,\,\,\,\,\,\,\,\,\,\,\,\,\,95 & \,\,\,\,\,\,\,\,\,\,\,\,\,\,\,5 & \,\,\,\,\,\,\,\,\,\,\,\,\,\,\,\,\,\,\,\,\,\,\,\,100 & \,\,\,\,\,\,\,\,\,\,\,\,\,\,\,\,\,\,\,\,\,\,\,\,0& \,\,\,\,\,\,\,\,\,\,\,\,\,\,\,\,\,\,\,\,\,\,\,\,\,\,\,\,\,\,\,\,\,\,\,\,\,\,99 & \,\,\,\,\,\,\,\,\,\,\,\,\,\,\,\,\,\,\,\,\,\,\,\,\,\,\,\,\,\,\,\,\,\,\,\,\,\,1 \\
Male & 26 & 76 & \,\,\,\,\,\,\,\,\,\,\,\,\,\,\,12 & \,\,\,\,\,\,\,\,\,\,\,\,\,\,\,90 & \,\,\,\,\,\,\,\,\,\,\,\,\,\,\,\,\,\,\,\,\,\,\,\,\,8 & \,\,\,\,\,\,\,\,\,\,\,\,\,\,\,\,\,\,\,\,\,\,\,\,94 & \,\,\,\,\,\,\,\,\,\,\,\,\,\,\,\,\,\,\,\,\,\,\,\,\,\,\,\,\,\,\,\,\,\,\,\,\,\,\,7 &  \,\,\,\,\,\,\,\,\,\,\,\,\,\,\,\,\,\,\,\,\,\,\,\,\,\,\,\,\,\,\,\,\,\,\,\,\,\,95 \\ \hline
Misclassification rate & \multicolumn{2}{c}{0.16} & \multicolumn{2}{c}{0.08} & \multicolumn{2}{c}{0.04} & \multicolumn{2}{c}{0.04} \\ \hline
Log-likelihood & \multicolumn{2}{c}{-} & \multicolumn{2}{c}{-2688} & \multicolumn{2}{c}{-2603} & \multicolumn{2}{c}{-2326} \\ \hline
\end{tabular}}
\caption{Comparison of the VCMM's steps' clustering performance on the subset of {\fontfamily{pcr}\selectfont AIS} data.}
\label{table:AIS-result-vcmm}
\end{table}

\begin{figure}[H]
	\centering
	\renewcommand{\xshiftNodes}{0.1*\linewidth}
	\renewcommand{\yshiftLabels}{.0cm}  
\begin{tikzpicture}	[every node/.style = VineNode, node distance =1.6cm,scale=0.75, transform shape]
\node (v5)   {LBM}		
node             (v2)         [right of = v5, xshift = \xshiftNodes] {Wt}
node             (v1)         [right of = v2, xshift = \xshiftNodes] {BMI}
node             (v3)         [below of = v1] {PBF}			
node             (v4)         [left of = v3, xshift =- \xshiftNodes] {WBC}
node[TreeLabels] (T1)        [below of = v5] {Males} ;
\draw[color=black] (v5) to node[draw=none,  font = \labelsize,fill = none, above, yshift = \yshiftLabels] {SG(10.70/0.91)} (v2);
\draw[color=black] (v2) to node[draw=none,  font = \labelsize,fill = none, above, yshift = \yshiftLabels] {N(0.83/0.63)} (v1);
\draw[color=black] (v1) to node[draw=none, font = \labelsize, fill = none, left, xshift = 0.5cm, yshift = \yshiftLabels] {BB1(0.69, 1.45/0.49)} (v3);
\draw[color=black] (v3) to node[draw=none, font = \labelsize, fill = none, below, yshift = \yshiftLabels] {F(2.59/0.27)} (v4);
\node (v5f)   [right of = v1, xshift=3*\xshiftTree]     {PBF}			
node             (v2f)         [right of = v5f, xshift = \xshiftNodes] {Wt}
node             (v1f)         [right of = v2f, xshift = \xshiftNodes] {LBM}
node             (v3f)         [below of = v2f] {BMI}			
node             (v4f)         [right of = v3f, xshift = \xshiftNodes] {WBC}
node[TreeLabels] (T2)        [below of =v5f] {Females};
\draw[color=black] (v5f) to node[draw=none,  font = \labelsize,fill = none, above, yshift = \yshiftLabels] {SBB8(4.07, 0.82/0.48)} (v2f);
\draw[color=black] (v2f) to node[draw=none,  font = \labelsize,fill = none, above, yshift = \yshiftLabels] {BB1(1.05, 2.81/0.77)} (v1f);
\draw[color=black] (v2f) to node[draw=none,font = \labelsize, fill = none,left, xshift = 0.5cm, yshift = \yshiftLabels] {N(0.87/0.67)} (v3f);
\draw[color=black] (v3f) to node[draw=none, font = \labelsize, fill = none, above, yshift = \yshiftLabels] {F(1.55/0.16)} (v4f);	
\end{tikzpicture}
\caption{The first tree level of the estimated vine copula model for females and males. A letter at an edge with numbers inside the parenthesis refers to its bivariate copula family with its estimated parameter(s)/Kendall's $\hat{\tau}$, where N: Gaussian, SG: Survival Gumbel, F: Frank, BB1: BB1, and SBB8: Survival BB8 copula.}
\label{fig:AIS-vine}
\end{figure}
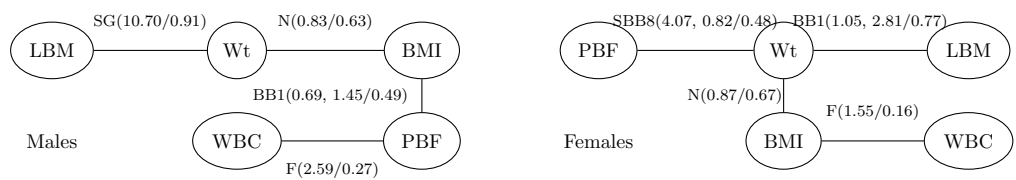

Figure \ref{fig:AIS-vine} illustrates the estimated vine copula models' first tree level by the final VCMM. The estimated vine tree structure is not the same for females and males. It is a path, i.e., D-vine, for males. The estimated pairwise dependence between pairs of the variables is non-Gaussian with diverse dependence strengths in females and males, except the pair of Wt and BMI. For instance, the pair of the variables, PBF and Wt, shows high strength, non-Gaussian dependence (Survival BB8 copula) in females. \ref{app-AIS} presents all estimated model components and parameters by the VCMM. In higher tree levels, the highest (absolute value) estimated Kendall's $\tau$ for females is $0.74$ and exists in the second tree, while $0.76$ in the third tree for males, i.e., high strength conditional dependence exists in the higher tree levels. The VCMM fits a log-logistic distribution for BMI in both clusters, while other variables' marginal distributions are different in males and females. The VCMM runs for $2.28$ minutes, and the ECM iterations stop after 15 iterations.

\begin{figure}[ht]
        \centering
        \begin{subfigure}[b]{0.15\textwidth}
            \centering
            \includegraphics[width=\textwidth]{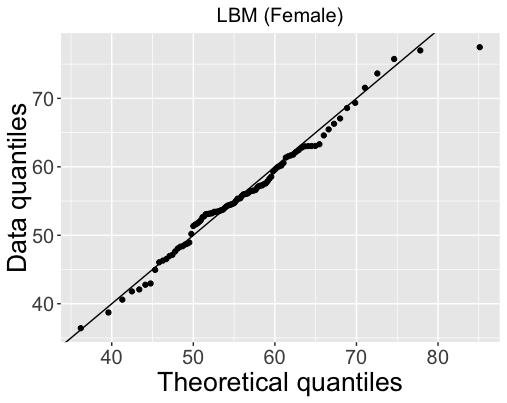}
        \end{subfigure}
        \begin{subfigure}[b]{0.15\textwidth}  
            \centering 
            \includegraphics[width=\textwidth]{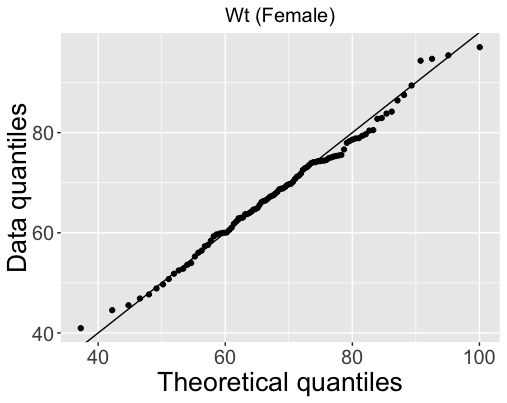}
        \end{subfigure}
        \begin{subfigure}[b]{0.15\textwidth}  
            \centering 
            \includegraphics[width=\textwidth]{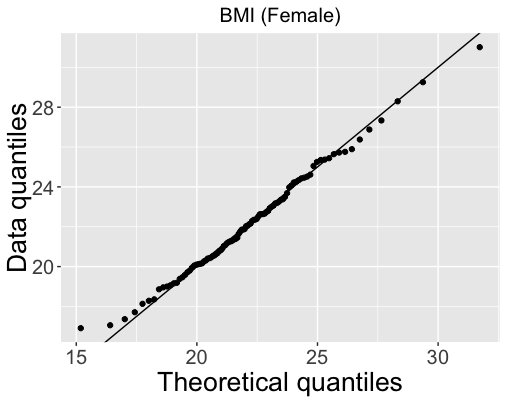}
        \end{subfigure}
        \begin{subfigure}[b]{0.15\textwidth}  
            \centering 
            \includegraphics[width=\textwidth]{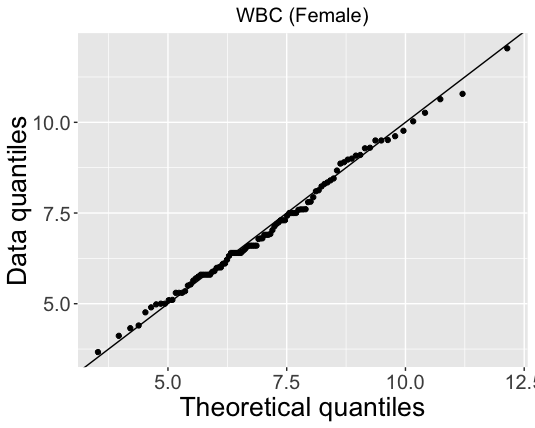}
        \end{subfigure}
        \begin{subfigure}[b]{0.15\textwidth}  
            \centering 
            \includegraphics[width=\textwidth]{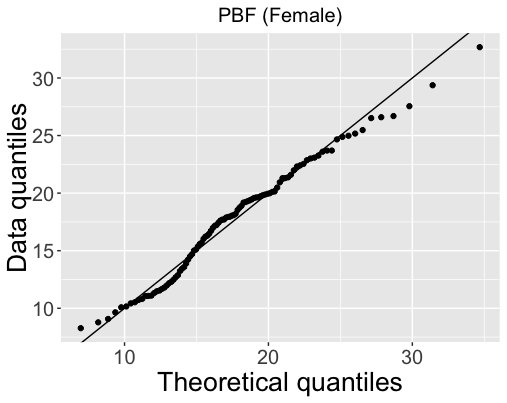}
        \end{subfigure}
        \vskip\baselineskip
        \begin{subfigure}[b]{0.15\textwidth}
            \centering
            \includegraphics[width=\textwidth]{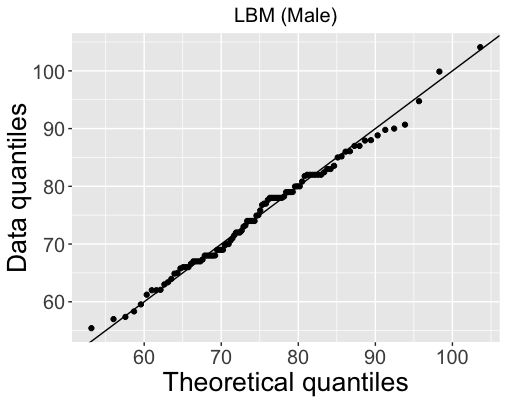}
        \end{subfigure}
        \begin{subfigure}[b]{0.15\textwidth}  
            \centering 
            \includegraphics[width=\textwidth]{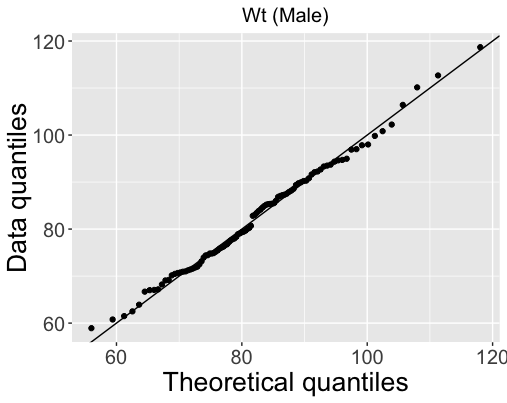}
        \end{subfigure}
        \begin{subfigure}[b]{0.15\textwidth}  
            \centering 
            \includegraphics[width=\textwidth]{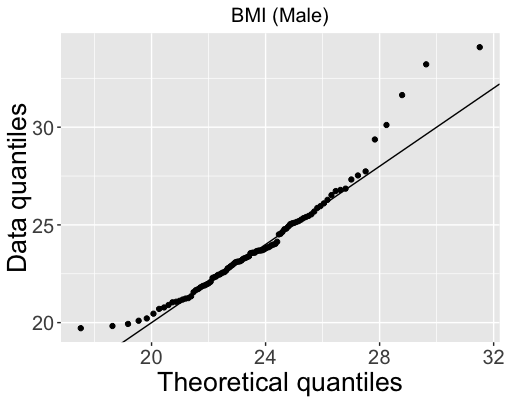}
        \end{subfigure}
        \begin{subfigure}[b]{0.15\textwidth}  
            \centering 
            \includegraphics[width=\textwidth]{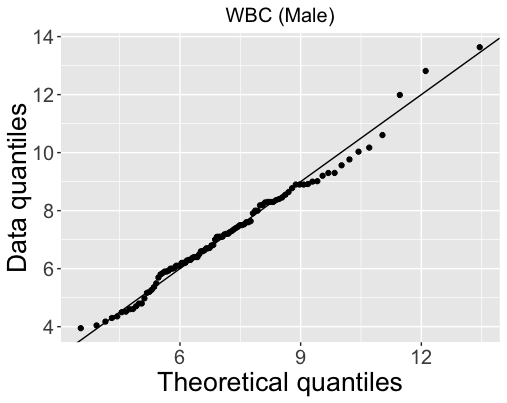}
        \end{subfigure}
        \begin{subfigure}[b]{0.15\textwidth}  
            \centering 
            \includegraphics[width=\textwidth]{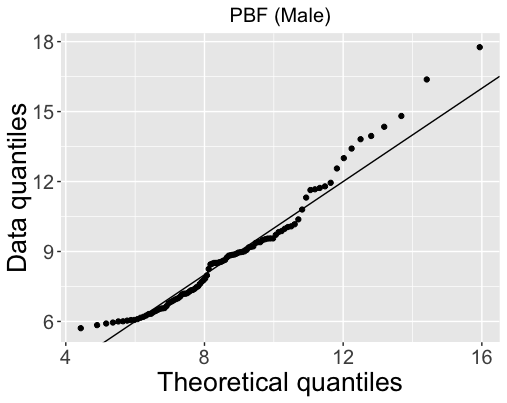}
        \end{subfigure}
        \caption{QQ plots of the variables in female (top) and male (bottom) clusters by the VCMM.}
\label{fig:AIS-qq}
\end{figure}

Figure \ref{fig:AIS-qq} shows the QQ plots of the variables in both fitted clusters. The fitted univariate marginal distributions by the VCMM are adequate for the female cluster. However, the margin selection could be improved for the variables BMI and PBF in the male cluster. For instance, adding univariate skew t distribution into the candidate list of univariate marginal distributions in line 9 of Algorithm \ref{vcmm-clustering-algo} could be considered. 
\subsection{Breast Cancer Wisconsin (Diagnostic)}\label{breast}
\begin{figure}[H]
\centering
\includegraphics[width=.45\textwidth]{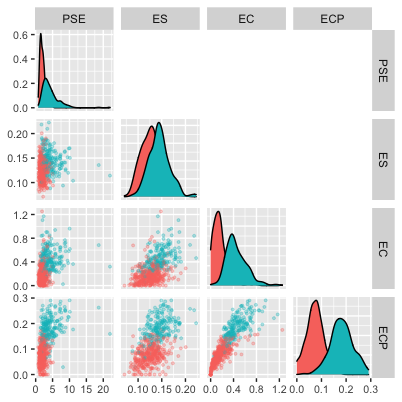}  \hspace{1cm}
\includegraphics[width=.45\textwidth]{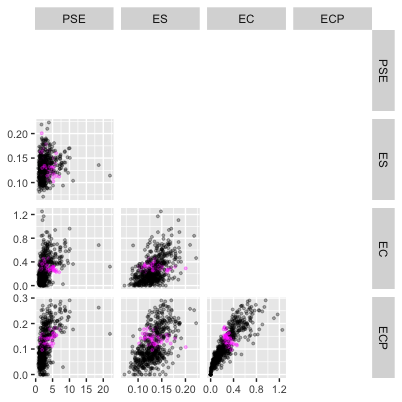} 
\caption{Pairwise scatter plots of the subset of {\fontfamily{pcr}\selectfont Breast Cancer Wisconsin (Diagnostic)} data. On the left panel, red/green points denote observations of benign/malignant. The right panel shows the VCMM clustering result, where magenta points show observations whose estimated posterior probability is smaller than 0.9 in their assigned cluster.}
\label{fig:breast}
\end{figure}

We now illustrate the effect of using a fixed vine tree structure for clusters in the VCMM and compare their performance with other model-based clustering algorithms.  We specify each cluster's vine tree structure as a star, i.e., C-vine, in the VCMM, selecting their root node with our approach explained in Section \ref{vcmm-selection}, and denote this model by VCMM(C-vine). We assess their performance on the Breast Cancer Wisconsin (Diagnostic) data obtained from the UCI Machine Learning data repository \citep{UCI}, where a digitized image of a fine needle aspirate (FNA) of a breast mass \citep{Mangasarian1995} is used to have ten features from 569 patients. The mean value, extreme value (mean of the three largest values), and standard error of each feature are calculated, returning 30 continuous variables. The data contains two types of diagnosis: benign (352 patients) and malignant (212 patients), enabling us to assess the misclassification rate of the binary classification for the algorithms. We limit this illustration to a subset of four variables: perimeter standard error (PSE), extreme smoothness (ES), extreme concavity (EC), and extreme concave points (ECP) shown on the left panel in Figure \ref{fig:breast}. The data has non-Gaussian variables like PSE. We fit two-component clustering models to the data, and k-means is not sensitive to seeds.
Table \ref{table:breast-result} shows that the VCMM provides a noticeably better clustering performance than other candidates. Even though its number of free parameters is higher than the mixture of multivariate normal distributions, not only does it achieve a lower BIC value, but the misclassification rate is also lower than theirs. The flexibility of the VCMM to model different dependence structures and marginal densities makes the difference. The selected vine tree structure by the VCMM is a path for the malignant cluster and a star with the root node of the variable ECP for the benign cluster. It runs for 1.07 minutes, and the ECM iterations stop after nine iterations. The right panel in Figure \ref{fig:breast} shows that the observations whose estimated posterior probability is smaller than 0.9 in their assigned cluster are at the border of the support regions of benign and malignant. \ref{app-breast} gives its estimated model components and the QQ plots of the fitted margins. 

Moreover, imposing a fixed vine structure in the VCMM, VCMM(C-vine), decreases the model power, where the selected root node is the variable ECP for both clusters. It appears to be the best vine copula mixture model regarding the BIC value. However, its misclassification rate is higher than the VCMM. Thus, we would need to construct a better model comparison criterion than the BIC value in the vine copula mixture model context in the future. The VCMM(C-vine) takes 7.14 minutes and 57 ECM iterations. 

\begin{table}[H]
\centering
\scalebox{0.75}{
\begin{tabular}{l  c  c  c   c  c  c c l}
Mixture model & VCMM & VCMM & Multivariate& Multivariate  & Multivariate & Multivariate\\
 &  & (C-vine) &normal & skew normal  & t & skew t\\
\hline
Misclassification rate & \textbf{0.10}&0.18&0.12&0.15&0.11&0.15\\
BIC&-3970& \textbf{-4014}&-3664&-3785 &-3727&-3858 \\
Number of free parameters &32&31&29&37&30&38\\
  \hline
\end{tabular}}
\caption{Comparison of model-based clustering algorithms' performance on the subset of {\fontfamily{pcr}\selectfont Breast Cancer Wisconsin (Diagnostic)} data. The best result in each row is highlighted.} 
\label{table:breast-result}
\end{table}
\subsection{Sachs Protein}\label{unsupervised}
\begin{figure}[ht]
\centering
\includegraphics[width=.45\textwidth]{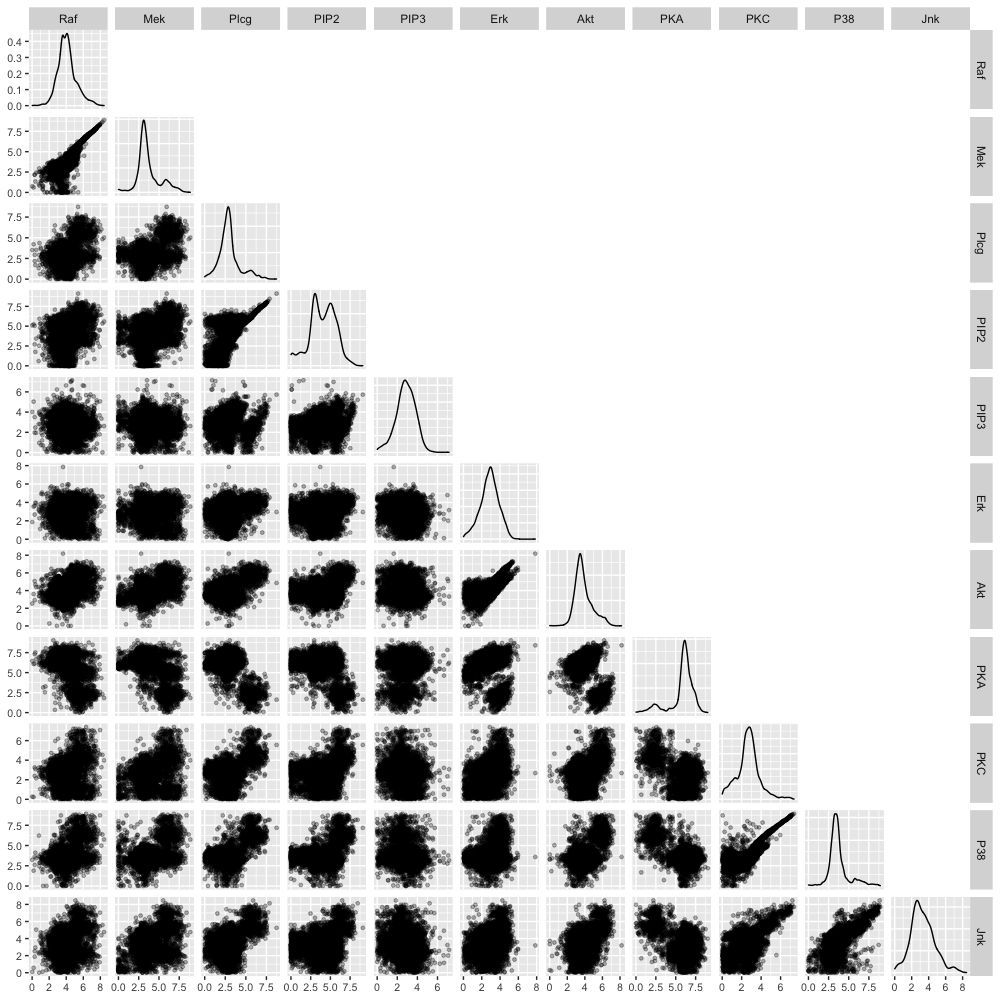} \hspace{1cm}
\includegraphics[width=.45\textwidth]{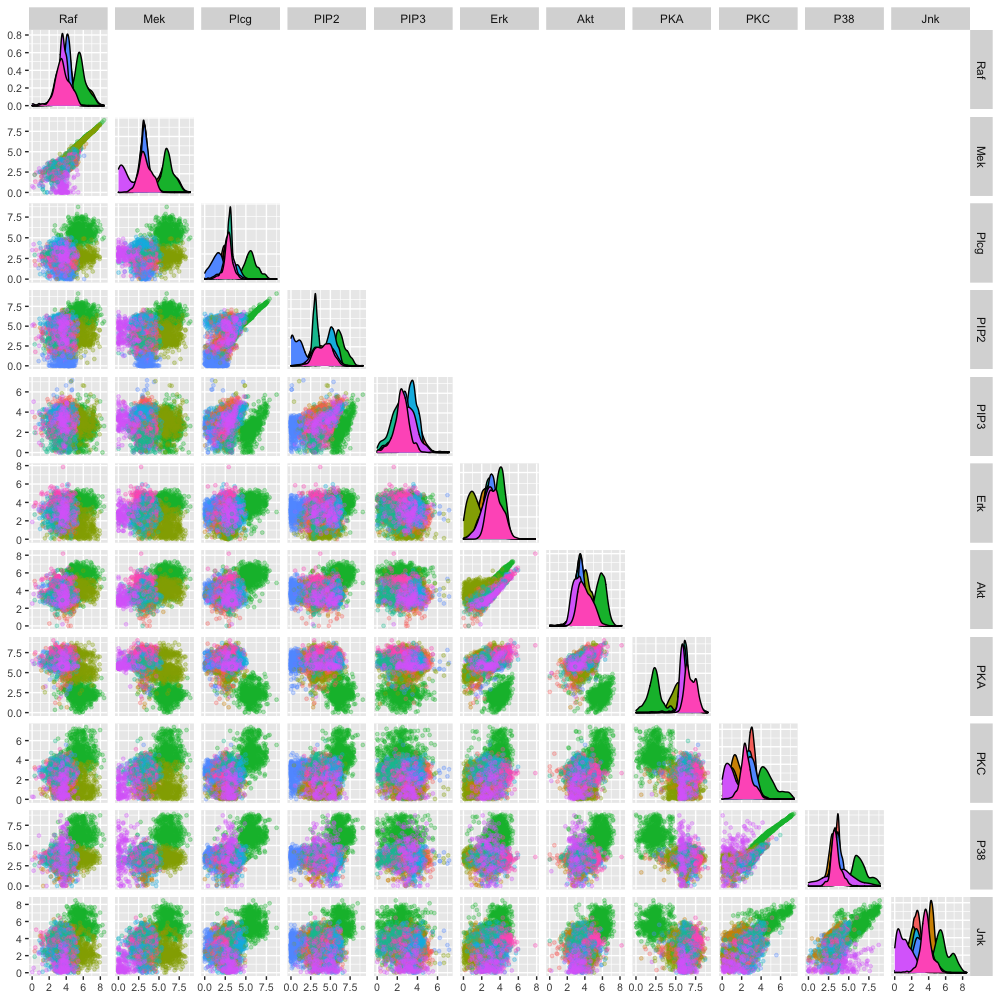}
\caption{Pairwise scatter plots of the {\fontfamily{pcr}\selectfont Sachs Protein} data (left) and its partition by the VCMM (right). The diagonal of the plots shows the corresponding variable's marginal density function in each cluster.}
\label{fig:sachs}
\end{figure}

We consider the Sachs Protein data analyzed by \cite{Sachs2005}. It consists of logarithmized levels of 11 phosphorylated proteins and phospholipids in individual cells, subjected to general and specific molecular interventions. The original goal is to learn the causal pathways linking a set of 11 proteins and compare them to the known links in the literature, thereby validating the important model tools in genetics, Bayesian networks. The data is continuous as shown in the left panel of Figure  \ref{fig:sachs}, where we work with 6161 observations from nine experiments ($b2camp$, $cd3cd28$, $cd3cd28 + aktinhib$, $cd3cd28 + g007$, $cd3cd28 + ly$, $cd3cd28 + psitect$, $cd3cd28 + u0126$, $cd3cd28icam2$, $pma$) after removing 1305 observations with zero values to avoid dealing with zero inflation. The standard approach in the literature is to assume that the data follows a multivariate Gaussian distribution, thereby formulating a Gaussian Bayesian network (GBN). However, Figure \ref{fig:sachs} shows that the data has non-Gaussian univariate distributions and different components. Accordingly, \cite{Zhang2017} works with the two-component Gaussian mixture copula Bayesian network and detects the underlying causal links better than the GBN.

\begin{figure}[H]
\begin{subfigure}[b]{0.45\linewidth}
\centering
\includegraphics[width=.7\textwidth]{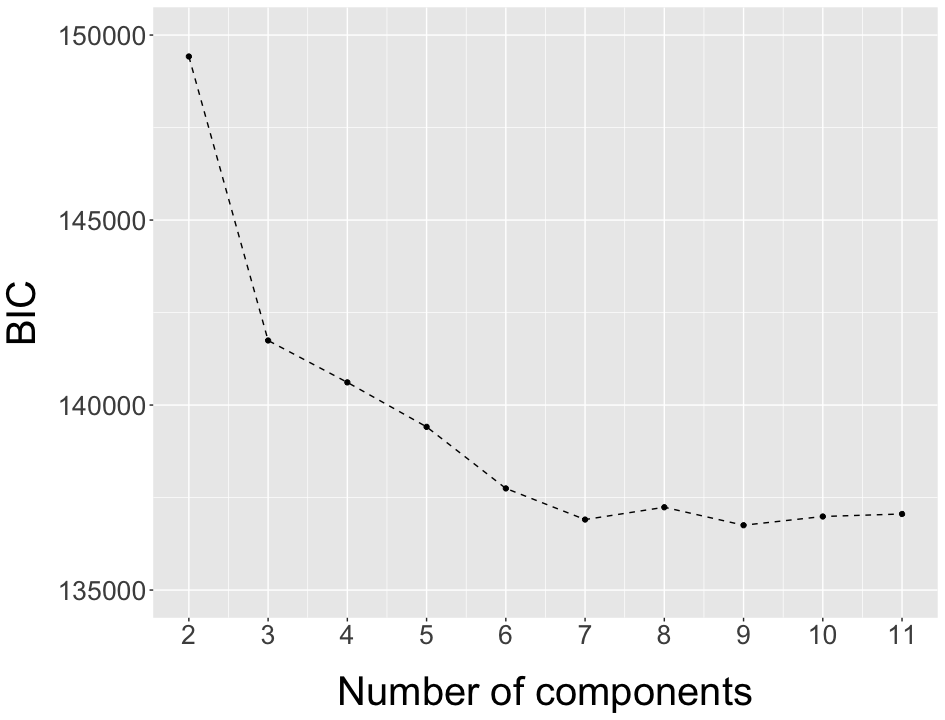} 
\caption{VCMM with k-means}
\end{subfigure}
\qquad
\begin{subfigure}[b]{0.45\linewidth}
\centering
\includegraphics[width=.7\textwidth]{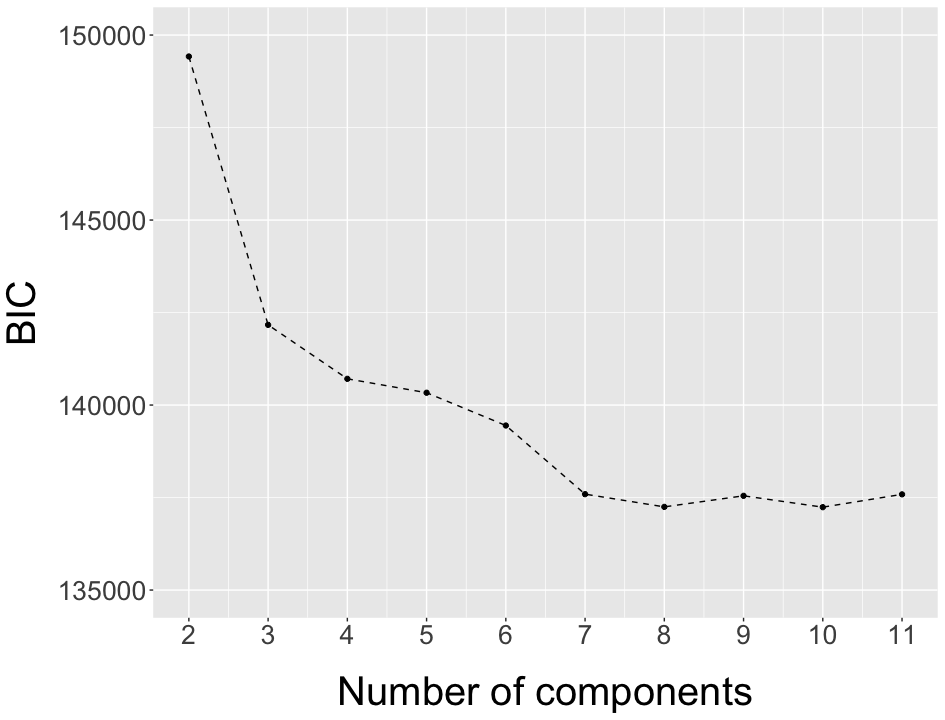}
\caption{VCMM with model-based hierarchical clustering}
\end{subfigure}
\caption{BIC values for the VCMM with different starting partitions and number of components on the {\fontfamily{pcr}\selectfont Sachs Protein} data.}
\label{fig:BIC-unsupervised}
\end{figure}

Even though bimodality exists for some variables, such as PIP2 and mek, and there are two obvious groups on some scatter plots like the one of Akt and PKA, it is unclear how many hidden components exist in the data. Therefore, following our suggestion in the previous sections, we fit the VCMM using a starting partition of k-means and model-based hierarchical clustering with two to eleven components. Figure \ref{fig:BIC-unsupervised} shows that using the initial partition from model-based hierarchical clustering in the VCMM suggests having ten components because it is the global minimum point of its plot. However, starting with the assignment of k-means points out nine components in the experiments, giving the lowest BIC value among the evaluated ones. Thus, we select it as our final VCMM model. Table \ref{table:unsupervised} shows that most observations of the experiments $b2camp$, $cd3cd28+g007$, $cd3cd28+psitect$, $cd3cd28+u0126$ are revealed as a cluster. The remaining five experiments' observations usually belong to the other four clusters. Their partition is given in the right panel of Figure \ref{fig:sachs}. The fitted univariate marginal distributions and bivariate copula families are mostly non-Gaussian in the clusters, showing the need for a non-Gaussian model. The fitted VCMM is available from the authors upon request.
\begin{table}[H]
\centering
\scalebox{0.6}{
\begin{tabular}{ccccccccccc}
\hline
Experiment& & $b2camp$ & $cd3cd28$ & $cd3cd28+aktinhib$ & $cd3cd28+g007$ & $cd3cd28+ly$ & $cd3cd28+psitect$ & $cd3cd28+u0126$ & $cd3cd28icam2$& $pma$ \\ 
  \hline
\multirow{9}{*}{Cluster}&1 &   0 & 245 & 297 &   0 & 298 &  62 &   0 &  14 &   5 \\ 
 & 2 &   0 & 201 &  36 &   0 & 145 &  37 &   4 &  26 &  33 \\ 
 & 3 &   0 &   3 &   2 &   0 &   0 &   1 & 500 &   0 &   1 \\ 
 & 4 &   0 &   0 &   0 & 695 &   0 &   0 &   0 &   0 &   0 \\ 
 & 5 &   3 & 143 &  36 &   0 & 125 &   6 &   2 & 246 & 310 \\ 
  &6 &   4 & 170 & 405 &   0 & 168 &   8 &   0 & 203 & 198 \\ 
&  7 &   0 &  23 &  39 &   0 &  42 & 473 &   1 &   8 &  25 \\ 
&  8 & 147 &  12 &  15 &   0 &   8 &   7 &   1 &  22 &  13 \\ 
 & 9 &   1 &  20 &  15 &   0 &  15 &  10 &   0 & 344 & 288 \\ 
   \hline
\end{tabular}}
\caption{Clustering result for the VCMM with nine components using the starting partition of k-means on the {\fontfamily{pcr}\selectfont Sachs Protein} data.}
\label{table:unsupervised}
\end{table}
\section{Conclusion}\label{conclusion}
We propose a vine copula mixture model that works with continuous data and fits all classes of vine tree structures. It uses parametric marginal distributions and pair copula families. It applies a wide range of pair copula families; thus, it accommodates diverse tail dependence and asymmetries within the components. Due to its parametric nature, it nicely interprets the structure of the data. Assuming the number of components in the data is known, we follow a data-driven approach for the remaining model selection problems. We work with the ECM algorithm for parameter estimation. With the proposed method, we formulate a new model-based clustering algorithm called VCMM. 

We evaluate the performance of the algorithm on simulated and real data. Our simulation studies illustrate that the vine copula based clustering has greater flexibility than various model-based clustering algorithms available in the literature and hence captures the non-Gaussian components hidden in the data better than others, especially when the data has heavy-tailed margins and tail dependence between pairs of variables. The real data analysis supports it and the better clustering assignment thanks to allowing all types of vine tree structures. Due to its flexibility in the formulation, it can also capture Gaussian components. Additionally, it can answer whether the dependence between a pair of variables changes with the variables’ different values or differs among the clusters. It is an appealing clustering approach since the era of big data comes with different data characteristics. 

To provide guidelines on the number of components to consider, we give a first analysis based on the BIC criterion. It turns out that the initial partition impacts the final model. Therefore, one possible future research direction is the selection of the number of components in the vine copula mixture models. It can be combined with the construction of a model comparison criterion. The ideas for sparse model selection in \cite{Nagler2019b} can be a starting point.

While the current algorithm has significant advancement in revealing non-elliptical components, further development would modify the proposed method to obtain parsimonious vine copula mixture models and then use them for clustering. On the one hand, the optimal truncation level can be studied, and, on the other hand, the parsimonious vine factor specification of \cite{Krupskii2013} can be considered. 

A potential drawback of the proposed method is the computational cost for high-dimensional data. This paper is an initial framework for using vine copulas with finite mixture models and clustering. Therefore, another future research direction is to handle variable selection and dimensionality reduction for vine copula based clustering. The traditional variable selection methods for other model-based clustering algorithms have to be reviewed and adjusted (e.g., \cite{Raftery2006}, \cite{Maugis2009}). Furthermore, the performance of the ECM algorithm's extensions, such as the expectation conditional maximization of either algorithm \citep{Liu1994}, can be analyzed for our framework. 

We show that different quick clustering methods can initialize the algorithm, and the final model can differ accordingly. Since running the algorithm until the stopping condition holds takes time with high-dimensional data, initialization approaches for the vine copula mixture models need to be further improved in the future. \cite{Scrucca2015} studies a similar problem with regard to the initial parameter values in multivariate normal mixture models. Initializing the algorithm could be further studied for harder situations with significant overlaps among the components and an unknown number of components.

The proposed method can be extended to deal with mixed discrete/continuous variables and missing data as other future research directions. The construction defined in \cite{Panagiotelis2012} and further studied in \cite{Panagiotelis2017} can be a starting point for the former. \cite{Wang2015} discusses how to handle missing data in multivariate skew t mixture models.

\section*{Acknowledgements}
This research has been supported by the German Research Foundation (DFG grant CZ 86/6-1). We are grateful to two anonymous referees and special issue editors for comments leading to a considerably improved contribution.
\appendix
\section{Abbreviation for univariate marginal distributions}\label{app-margin}
\noindent
$lnorm(\mu,\sigma)$:log-normal distribution with mean/standard deviation parameters $\mu$/$\sigma$ on the logarithmic scale, 
\newline
$exp(\lambda)$: exponential distribution with rate parameter $\lambda$, 
\newline
$llogis(\alpha, \beta)$: log-logistic distribution with shape parameter $\alpha$ and scale parameter $\beta$, 
\newline
$logis(l, s)$: logistic distribution with location parameter $l$ and scale parameter $s$,  
\newline
$\Gamma(\alpha,\beta)$: gamma distribution with shape parameter $\alpha$ and rate parameter $\beta$,  
\newline
$\mathcal{N}(\mu,\sigma)$: normal distribution with mean parameter $\mu$ and standard deviation parameter $\sigma$.
\newline
$t_3(\mu,\sigma)$: Student's t distribution with mean/standard deviation parameters $\mu$/$\sigma$, and degrees of freedom 3.
\section{Overview of the VCMM}\label{app-flow}

\begin{figure}[H]
	\centering
	\renewcommand{\xshiftNodes}{0.1*\linewidth}
	\renewcommand{\yshiftLabels}{.0cm}  
	\renewcommand{\labelsize}{\scriptsize} 
  \begin{tikzpicture}[node distance =1.7cm,scale=0.75, transform shape,
  disc/.style = {shape=cylinder, draw, shape aspect=0.3,shape border rotate=90,
  text width=17mm, align=center},
  mdl/.style = {shape=ellipse, draw, shape aspect=0.3,shape border rotate=90,
  text width=17mm, align=center}, 
  alg/.style = {shape=rectangle, draw, shape aspect=0.3,shape border rotate=90,
  text width=17mm, align=center}]
\node (v1) [disc] {Data\\ on $x$-scale}
node (v2) [mdl, right of = v1, xshift = 2*\xshiftNodes]  {Initial \\ partition \\ of the data}
node (v3) [alg, right of = v2, xshift = \xshiftNodes]  {Initial margin \\ selection for each cluster}
node (v4) [disc, right of = v3, xshift = 2*\xshiftNodes] {Data\\ on $u$-scale for each cluster}
node (v5) [alg, right of = v4, xshift = \xshiftNodes] {Markov tree \\ selection for each cluster}
node (v6) [alg, below of = v5, yshift = -1.4*\xshiftNodes] {Parameter estimation via ECM for each cluster}
node (v7) [mdl, left of = v6,  xshift = -2*\xshiftNodes]  {Temporary \\ partition \\ of the data}
node (v8) [alg, left of = v7,  xshift = -\xshiftNodes]  {Final margin \\ selection for each cluster}
node (v9) [disc, left of = v8, xshift = -0.8*\xshiftNodes] {Data\\ on $u$-scale for each cluster}
node (v10) [alg, left of = v9, xshift = -0.8*\xshiftNodes] {Vine tree \\ selection for each cluster}
node (v11) [mdl, left of = v10,  xshift = -1.4*\xshiftNodes] {Final \\ partition \\ of the data};
\draw[->, color=black] (v1) to node[draw=none,  font = \labelsize,fill = none, above, yshift = \yshiftLabels] {$k$ component} (v2);
\draw[->, color=black] (v1) to node[draw=none,  font = \labelsize,fill = none, below, yshift = \yshiftLabels] { fast clustering} (v2);
\draw[->, color=black] (v2) to node[draw=none,  font = \labelsize,fill = none, above, yshift = \yshiftLabels] {via} (v3);
\draw[->, color=black] (v2) to node[draw=none,  font = \labelsize,fill = none, below, yshift = \yshiftLabels] {BIC} (v3);
\draw[->, color=black] (v3) to node[draw=none,  font = \labelsize,fill = none, above, yshift = \yshiftLabels] {Probability integral} (v4);
\draw[->, color=black] (v3) to node[draw=none,  font = \labelsize,fill = none, below, yshift = \yshiftLabels] {transformation (PIT)} (v4);
\draw[->, color=black] (v4) to node[draw=none,  font = \labelsize,fill = none, above, yshift = \yshiftLabels] {via} (v5);
\draw[->, color=black] (v4) to node[draw=none,  font = \labelsize,fill = none, below, yshift = \yshiftLabels] {AIC} (v5);
\draw[->, color=black] (v5) to node[draw=none,  font = \labelsize,fill = none, above, xshift = 1cm] {Keeping initial Markov} (v6);
\draw[->, color=black] (v5) to node[draw=none,  font = \labelsize,fill = none, below, xshift = 1cm] {copula models fixed} (v6);
\draw[->, color=black] (v6) to node[draw=none,  font = \labelsize,fill = none, above, yshift = \yshiftLabels] {If the termination} (v7);
\draw[->, color=black] (v6) to node[draw=none,  font = \labelsize,fill = none, below, yshift = \yshiftLabels] {criterion holds} (v7);
\draw[->, color=black] (v7) to node[draw=none,  font = \labelsize,fill = none, above, yshift = \yshiftLabels] {via} (v8);
\draw[->, color=black] (v7) to node[draw=none,  font = \labelsize,fill = none, below, yshift = \yshiftLabels] {BIC} (v8);
\draw[->, color=black] (v8) to node[draw=none,  font = \labelsize,fill = none, above, yshift = \yshiftLabels] {PIT} (v9);
\draw[->, color=black] (v9) to node[draw=none,  font = \labelsize,fill = none, above, yshift = \yshiftLabels] {via} (v10);
\draw[->, color=black] (v9) to node[draw=none,  font = \labelsize,fill = none, below, yshift = \yshiftLabels] {AIC} (v10);
\draw[->, color=black] (v10) to node[draw=none,  font = \labelsize,fill = none, above, yshift = \yshiftLabels] {via} (v11);
\draw[->, color=black] (v10) to node[draw=none,  align=center, font = \labelsize,fill = none, below, yshift = \yshiftLabels] {$final$\\$posterior$ \\  $probabilities$} (v11);
    \end{tikzpicture}
\end{figure}
\section{Specification of the artificial data generation in Section \ref{vcmm-sim-4}}\label{app-sim}
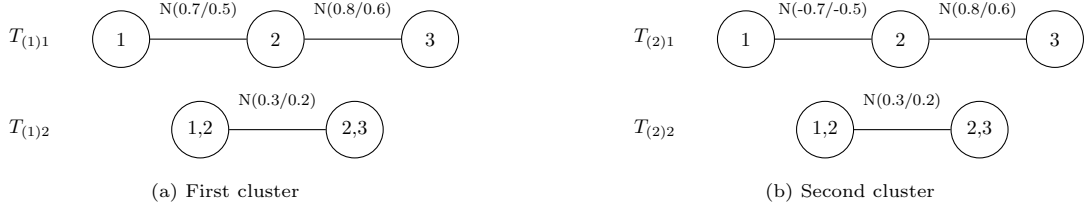
\begin{figure}[H]
\begin{subfigure}[b]{0.45\linewidth}
	\centering
	\renewcommand{\xshiftNodes}{0.15*\linewidth}
	\renewcommand{\yshiftLabels}{.0cm}  
\begin{tikzpicture}	[every node/.style = VineNode, node distance =1.6cm,scale=0.75, transform shape]
\node[TreeLabels] (T1) {$T_{(1)1}$}
node             (v1)         [right of = T1] {1}			
node             (v2)         [right of = v1, xshift = \xshiftNodes] {2}
node             (v3)         [right of = v2, xshift = \xshiftNodes] {3};
\draw[color=black] (v1) to node[draw=none,  font = \labelsize,fill = none, above, yshift = \yshiftLabels] {N(0.7/0.5)} (v2);
\draw[color=black] (v2) to node[draw=none,  font = \labelsize,fill = none, above, yshift = \yshiftLabels] {N(0.8/0.6)} (v3);
\node[TreeLabels] (T2)      [below of = T1] {$T_{(1)2}$}
node             (v13)         [right of = T2, xshift = 1.25*\xshiftNodes] {1,2}
node             (v23)         [right of = v13, xshift =\xshiftNodes] {2,3};
\draw[color=black] (v13) to node[draw=none,  font = \labelsize,fill = none, above, yshift = \yshiftLabels] {N(0.3/0.2)} (v23);
\end{tikzpicture}
\caption{First cluster}
\end{subfigure}
\qquad
\begin{subfigure}[b]{0.45\linewidth}
	\centering
	\renewcommand{\xshiftNodes}{0.15*\linewidth}
	\renewcommand{\yshiftLabels}{.0cm}  
\begin{tikzpicture}	[every node/.style = VineNode, node distance =1.6cm,scale=0.75, transform shape]
\node[TreeLabels] (T1)   {$T_{(2)1}$}
node             (v1)         [right of = T1] {1}			
node             (v2)         [right of = v1, xshift = \xshiftNodes] {2}
node             (v3)         [right of = v2, xshift = \xshiftNodes] {3};
\draw[color=black] (v1) to node[draw=none, font = \labelsize, fill = none, above, yshift = \yshiftLabels] {N(-0.7/-0.5)} (v2);
\draw[color=black] (v2) to node[draw=none,  font = \labelsize,fill = none, above, yshift = \yshiftLabels] {N(0.8/0.6)} (v3);
\node[TreeLabels] (T2)      [below of = T1] {$T_{(2)2}$}
node             (v13)         [right of = T2, xshift =1.25*\xshiftNodes] {1,2}
node             (v23)         [right of = v13, xshift =\xshiftNodes] {2,3};
\draw[color=black] (v13) to node[draw=none,  font = \labelsize,fill = none, above, yshift = \yshiftLabels] {N(0.3/0.2)} (v23);
\end{tikzpicture}
\caption{Second cluster}
\end{subfigure}
\vspace{-4mm}
\caption{Vine tree structure of simulated data with two clusters. A letter at an edge refers to its bivariate copula family, where N: Gaussian copula. The true parameter value and corresponding Kendall's $\tau$ of the pair copula are given inside the parenthesis (parameter/Kendall's $\tau$).}
\label{fig:vcmm-case-3}
\end{figure}
\begin{table}[H]
\centering
\scalebox{0.75}{
\begin{tabular}{l  l l  l l  l}
\hline
 $F_{1(1)}(\bm{\gamma}_{1(1)})$ &  $F_{2(1)}(\gamma_{2(1)})$&  $F_{3(1)}(\gamma_{3(1)})$&  $F_{1(2)}(\bm{\gamma}_{1(2)})$&  $F_{2(2)}(\bm{\gamma}_{2(2)})$&  $F_{3(2)}(\gamma_{3(2)})$ \\ 
 $\mathcal{N}(0,2)$ & $\mathcal{N}(1,2)$& $\mathcal{N}(1,2)$& $\mathcal{N}(0,2)$ & $\mathcal{N}(1,2)$&  $\mathcal{N}(-2,2)$\\ 
  \hline
\end{tabular}}
\caption{Univariate marginal distributions and associated parameters of each cluster. They are defined in \ref{app-margin}.}
\label{table:vcmm-case-3}
\end{table}
\section{Estimated model components and parameters of the {\fontfamily{pcr}\selectfont AIS} data by the VCMM}\label{app-AIS}
The variable encoding is given as follows: 1: LBM, 2: Wt, 3: BMI, 4: WBC, and 5: PBF. The cluster index $(1)$ refers to females, whereas the index $(2)$ denotes males.
\begin{table}[H]
\centering
\scalebox{0.75}{
\begin{tabular}{l l l l  l l  l}
\hline
 j & $\hat{F}_{1(j)}(\hat{\bm{\gamma}}_{1(j)})$ &  $\hat{F}_{2(j)}(\hat{\bm{\gamma}}_{2(j)})$&  $\hat{F}_{3(j)}(\hat{\bm{\gamma}}_{3(j)})$&  $\hat{F}_{4(j)}(\hat{\bm{\gamma}}_{4(j)})$&  $\hat{F}_{5(j)}(\hat{\bm{\gamma}}_{5(j)})$&  $\hat{\pi}_j$ \\ 
 1 & $llogis(12.4, 55.6)$  & $\mathcal{N}(68.6, 12.1)$ &  $llogis(14.4, 22.0)$ &  $\Gamma(18.0, 2.5)$   &  $\Gamma(10.8, 0.6)$   &  $0.54$\\ 
2 &  $lnorm(4.3, 0.1)$ &  $lnorm(4.4, 0.1)$ &  $llogis(17.9, 23.5)$ &  $lnorm(1.9, 0.3)$  & $lnorm(2.1,0.2)$ &  $0.46$\\ 
  \hline
\end{tabular}}
\caption{Estimated mixture weight, marginal distributions and parameters of females and males. The marginal distributions are defined in \ref{app-margin}.} 
\label{table:AIS-margin}
\vspace{-1cm}
\end{table}
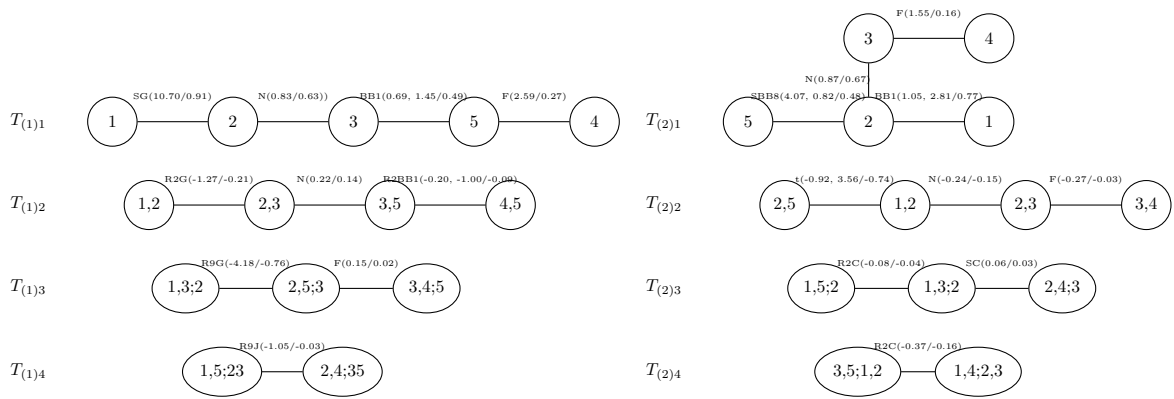
\begin{figure}[H]
\begin{subfigure}[b]{0.45\linewidth}
	\centering
	\renewcommand{\xshiftNodes}{0.1*\linewidth}
	\renewcommand{\yshiftLabels}{.0cm}  
	\renewcommand{\labelsize}{\tiny} 
\begin{tikzpicture}	[every node/.style = VineNode, node distance =1.7cm,scale=0.65, transform shape]
\node[TreeLabels] (T1)       {$T_{(1)1}$}
node (v5)  [right of = T1] {1}		
node             (v2)         [right of = v5, xshift = \xshiftNodes] {2}
node             (v1)         [right of = v2, xshift = \xshiftNodes] {3}
node             (v3)         [right of = v1, xshift = \xshiftNodes] {5}			
node             (v4)         [right of = v3, xshift = \xshiftNodes] {4};
\draw[color=black] (v5) to node[draw=none,  font = \labelsize,fill = none, above, yshift = \yshiftLabels] {SG(10.70/0.91)} (v2);
\draw[color=black] (v2) to node[draw=none,  font = \labelsize,fill = none, above, yshift = \yshiftLabels] {N(0.83/0.63))} (v1);
\draw[color=black] (v1) to node[draw=none, font = \labelsize, fill = none, above, yshift = \yshiftLabels] {BB1(0.69, 1.45/0.49)} (v3);
\draw[color=black] (v3) to node[draw=none, font = \labelsize, fill = none, above, yshift = \yshiftLabels] {F(2.59/0.27)} (v4);
\node[TreeLabels] (T2)      [below of = T1] {$T_{(1)2}$}
node             (v25)         [right of = T2, xshift =\xshiftNodes] {1,2}
node             (v12)         [right of = v25, xshift =\xshiftNodes] {2,3}
node             (v13)         [right of = v12, xshift =\xshiftNodes] {3,5}
node             (v34)         [right of = v13, xshift =\xshiftNodes] {4,5};
\draw[color=black] (v25) to node[draw=none, font = \labelsize,fill = none, above, yshift = \yshiftLabels] {R2G(-1.27/-0.21)} (v12);
\draw[color=black] (v12) to node[draw=none,  font = \labelsize,fill = none, above, yshift = \yshiftLabels] {N(0.22/0.14)} (v13);
\draw[color=black] (v13) to node[draw=none,  font = \labelsize,fill = none, above, yshift = \yshiftLabels] {R2BB1(-0.20, -1.00/-0.09)} (v34);
\node[TreeLabels] (T3)      [below of = T2] {$T_{(1)3}$}
node             (v152)         [right of = T3, xshift =2*\xshiftNodes] {1,3;2}
node             (v231)         [right of = v152, xshift =\xshiftNodes] {2,5;3}
node             (v143)         [right of = v231, xshift =\xshiftNodes] {3,4;5};
\draw[color=black] (v152) to node[draw=none,  font = \labelsize,fill = none, above, yshift = \yshiftLabels] {R9G(-4.18/-0.76)} (v231);
\draw[color=black] (v231) to node[draw=none,  font = \labelsize,fill = none, above, yshift = \yshiftLabels] {F(0.15/0.02)} (v143);
\node[TreeLabels] (T4)      [below of = T3] {$T_{(1)4}$}
node             (v3512)         [right of = T4, xshift =3*\xshiftNodes] {1,5;23}
node             (v2413)         [right of = v3512, xshift =\xshiftNodes] {2,4;35};
\draw[color=black] (v3512) to node[draw=none, font = \labelsize, fill = none, above, yshift = \yshiftLabels] {R9J(-1.05/-0.03)} (v2413);
\end{tikzpicture}
\end{subfigure}
\qquad
\begin{subfigure}[b]{0.45\linewidth}
	\centering
	\renewcommand{\xshiftNodes}{0.1*\linewidth}
	\renewcommand{\yshiftLabels}{.0cm}  
	\renewcommand{\labelsize}{\tiny} 
\begin{tikzpicture}	[every node/.style = VineNode, node distance =1.7cm,scale=0.65, transform shape]
\node[TreeLabels] (T1)       {$T_{(2)1}$}
node (v5f)   [right of = T1]     {5}			
node             (v2f)         [right of = v5f, xshift = \xshiftNodes] {2}
node             (v1f)         [right of = v2f, xshift = \xshiftNodes] {1}
node             (v3f)         [above of = v2f] {3}			
node             (v4f)         [right of = v3f, xshift = \xshiftNodes] {4};
\draw[color=black] (v5f) to node[draw=none,  font = \labelsize,fill = none, above, yshift = \yshiftLabels] {SBB8(4.07, 0.82/0.48)} (v2f);
\draw[color=black] (v2f) to node[draw=none,  font = \labelsize,fill = none, above, yshift = \yshiftLabels] {BB1(1.05, 2.81/0.77)} (v1f);
\draw[color=black] (v2f) to node[draw=none,font = \labelsize, fill = none,left, xshift = 0.5cm, yshift = \yshiftLabels] {N(0.87/0.67)} (v3f);
\draw[color=black] (v3f) to node[draw=none, font = \labelsize, fill = none, above, yshift = \yshiftLabels] {F(1.55/0.16)} (v4f);	
\node[TreeLabels] (T2)      [below of = T1] {$T_{(2)2}$}
node             (v25)         [right of = T2, xshift =\xshiftNodes] {2,5}
node             (v12)         [right of = v25, xshift =\xshiftNodes] {1,2}
node             (v13)         [right of = v12, xshift =\xshiftNodes] {2,3}
node             (v34)         [right of = v13, xshift =\xshiftNodes] {3,4};
\draw[color=black] (v25) to node[draw=none, font = \labelsize,fill = none, above, yshift = \yshiftLabels] {t(-0.92, 3.56/-0.74)} (v12);
\draw[color=black] (v12) to node[draw=none,  font = \labelsize,fill = none, above, yshift = \yshiftLabels] {N(-0.24/-0.15)} (v13);
\draw[color=black] (v13) to node[draw=none,  font = \labelsize,fill = none, above, yshift = \yshiftLabels] {F(-0.27/-0.03)} (v34);
\node[TreeLabels] (T3)      [below of = T2] {$T_{(2)3}$}
node             (v152)         [right of = T3, xshift =2*\xshiftNodes] {1,5;2}
node             (v231)         [right of = v152, xshift =\xshiftNodes] {1,3;2}
node             (v143)         [right of = v231, xshift =\xshiftNodes] {2,4;3};
\draw[color=black] (v152) to node[draw=none,  font = \labelsize,fill = none, above, yshift = \yshiftLabels] {R2C(-0.08/-0.04)} (v231);
\draw[color=black] (v231) to node[draw=none,  font = \labelsize,fill = none, above, yshift = \yshiftLabels] {SC(0.06/0.03)} (v143);
\node[TreeLabels] (T4)      [below of = T3] {$T_{(2)4}$}
node             (v3512)         [right of = T4, xshift =3*\xshiftNodes] {3,5;1,2}
node             (v2413)         [right of = v3512, xshift =\xshiftNodes] {1,4;2,3};
\draw[color=black] (v3512) to node[draw=none, font = \labelsize, fill = none, above, yshift = \yshiftLabels] {R2C(-0.37/-0.16)} (v2413);
\end{tikzpicture}
\end{subfigure}
\caption{Estimated vine copula models for females and males. A letter at an edge with numbers inside the parenthesis refers to its bivariate copula family with its parameter(s)/Kendall's $\hat{\tau}$, where t: t, SC: Survival Clayton, R2C: Rotated Clayton 270 degrees, R9G: Rotated Gumbel 90 degrees, R2G: Rotated Gumbel 270 degrees,  R9J: Rotated Joe 90 degrees,  R2BB1: Rotated BB1 270 degrees copula. Others are defined in Figure \ref{fig:AIS-vine}.}
\end{figure}
\section{Estimated model components and parameters of the {\fontfamily{pcr}\selectfont Breast Cancer Wisconsin (Diagnostic)} data by the VCMM}\label{app-breast}
The variable encoding is 1: perimeter standard error, 2: extreme smoothness, 3: extreme concavity, and 4: extreme concave points. The cluster encoding is $(1)$: malignant and $(2)$: benign.
\begin{table}[H]
\centering
\scalebox{0.75}{
\begin{tabular}{l l l l  l  l}
\hline
 j & $\hat{F}_{1(j)}(\hat{\bm{\gamma}}_{1(j)})$ &  $\hat{F}_{2(j)}(\hat{\bm{\gamma}}_{2(j)})$&  $\hat{F}_{3(j)}(\hat{\bm{\gamma}}_{3(j)})$&  $\hat{F}_{4(j)}(\hat{\bm{\gamma}}_{4(j)})$&  $\hat{\pi}_j$ \\ 
 1  & $lnorm(1.3, 0.5)$ &  $llogis(13.4, 0.1)$& $lnorm(-0.8, 0.3)$  &  $\Gamma(22.4, 118.6)$  &  $0.35$\\ 
2  &  $lnorm(0.7, 0.4)$ &  $\Gamma(43.2, 349.6)$ &  $\mathcal{N}(0.15, 0.1)$ &  $\mathcal{N}(0.07, 0.03)$ &  $0.65$\\ 
  \hline
\end{tabular}}
\caption{Estimated mixture weight, marginal distributions and parameters of benign and malignant. The marginal distributions are defined in \ref{app-margin}.} 
\label{table:breast-margin}
\vspace{-4mm}
\end{table}
\begin{figure}[H]
\begin{subfigure}[b]{0.45\linewidth}
	\centering
	\renewcommand{\xshiftNodes}{0.1*\linewidth}
	\renewcommand{\yshiftLabels}{.0cm}  
	\renewcommand{\labelsize}{\tiny} 
\begin{tikzpicture}	[every node/.style = VineNode, node distance =2cm,scale=0.7, transform shape]
\node[TreeLabels] (T1)       {$T_{(1)1}$}
node             (v1)         [right of = T1] {2}			
node             (v2)         [right of = v1, xshift = \xshiftNodes] {1}
node             (v3)         [right of = v2, xshift = \xshiftNodes] {4}
node             (v4)         [right of = v3, xshift= \xshiftNodes] {3};
\draw[color=black] (v1) to node[draw=none, font = \labelsize, fill = none, above, yshift = \yshiftLabels] {N(-0.27/-0.17)} (v2);
\draw[color=black] (v2) to node[draw=none, font = \labelsize, fill = none, above, yshift = \yshiftLabels] {N(0.42/0.27)} (v3);
\draw[color=black] (v3) to node[draw=none,  font = \labelsize,fill = none, above, yshift = \yshiftLabels] {BB8(6, 0.48/0.35)} (v4);
\end{tikzpicture}
\end{subfigure}
\qquad
\begin{subfigure}[b]{0.45\linewidth}
	\centering
	\renewcommand{\xshiftNodes}{0.1*\linewidth}
	\renewcommand{\yshiftLabels}{.0cm}  
	\renewcommand{\labelsize}{\tiny} 
	\begin{tikzpicture}	[every node/.style = VineNode, node distance =2cm,scale=0.7, transform shape]
\node[TreeLabels] (T1)       {$T_{(2)1}$}
node (v3)     [right of = T1] {3}		
node             (v4)         [right of = v3, xshift = \xshiftNodes] {4}
node             (v2)         [above of = v4,  yshift = 0.01*\linewidth] {2}
node             (v1)         [right of = v4, xshift = \xshiftNodes] {1};
\draw[color=black] (v4) to node[draw=none,  font = \labelsize,fill = none, left, xshift = 0.3cm, yshift = \yshiftLabels] {C(0.44/0.18)} (v2);
\draw[color=black] (v3) to node[draw=none, font = \labelsize, fill = none, above,  yshift = \yshiftLabels] {SBB8(3.90, 0.96/0.57)} (v4);
\draw[color=black] (v4) to node[draw=none, font = \labelsize, fill = none, above,  yshift = \yshiftLabels] {J(1.29/0.14)} (v1);
\end{tikzpicture}
\end{subfigure}
\caption{The first tree level of the estimated vine copula models for benign and malignant. A letter at an edge with numbers inside the parenthesis refers to its bivariate copula family with its parameter(s)/Kendall's $\hat{\tau}$, where C: Clayton, J: Joe and BB8: BB8 copula. Others are defined in Figure \ref{fig:AIS-vine}.}
\label{fig:breast-vine}
\vspace{-4mm}
\end{figure}
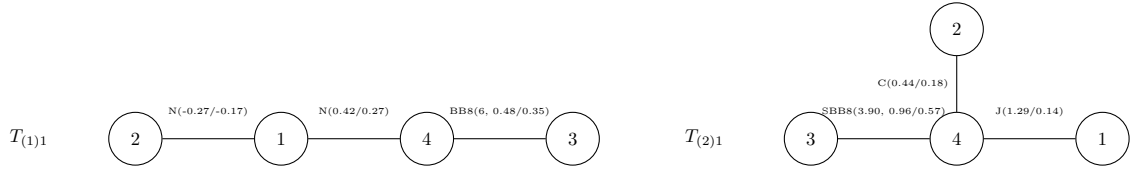
 \begin{figure}[ht]
        \centering
        \begin{subfigure}[b]{0.15\textwidth}
            \centering
            \includegraphics[width=\textwidth]{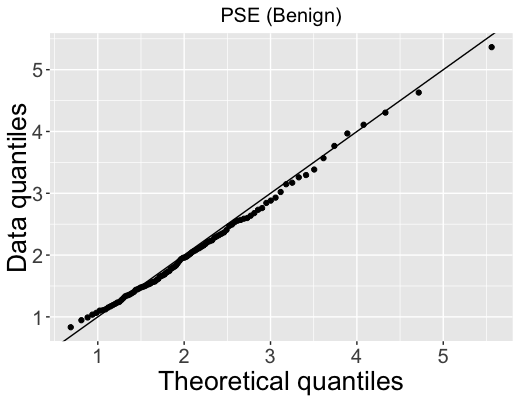}
        \end{subfigure}
        \begin{subfigure}[b]{0.15\textwidth}  
            \centering 
            \includegraphics[width=\textwidth]{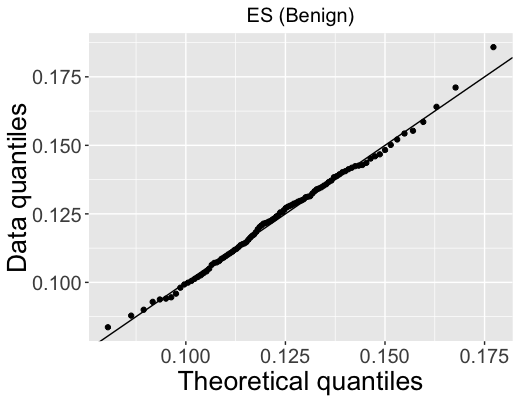}
        \end{subfigure}
        \begin{subfigure}[b]{0.15\textwidth}  
            \centering 
            \includegraphics[width=\textwidth]{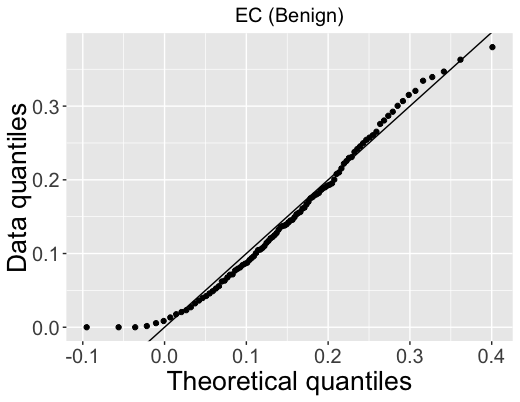}
        \end{subfigure}
        \begin{subfigure}[b]{0.15\textwidth}  
            \centering 
            \includegraphics[width=\textwidth]{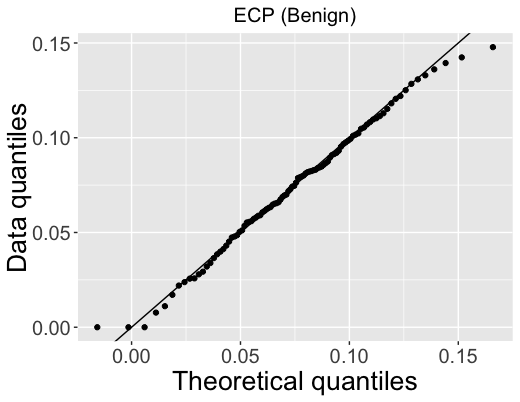}
        \end{subfigure}
        \vskip\baselineskip
        \begin{subfigure}[b]{0.15\textwidth}
            \centering
            \includegraphics[width=\textwidth]{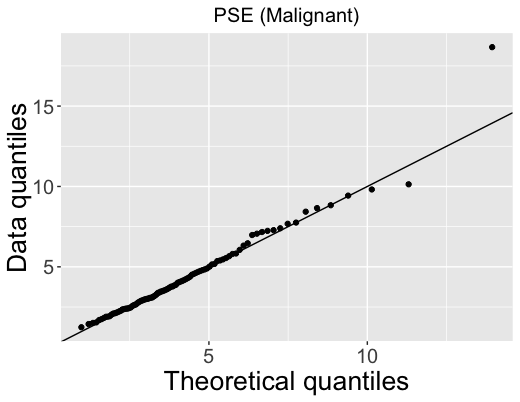}
        \end{subfigure}
        \begin{subfigure}[b]{0.15\textwidth}  
            \centering 
            \includegraphics[width=\textwidth]{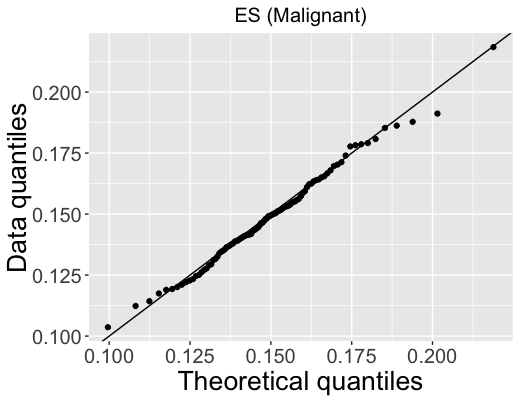}
        \end{subfigure}
        \begin{subfigure}[b]{0.15\textwidth}  
            \centering 
            \includegraphics[width=\textwidth]{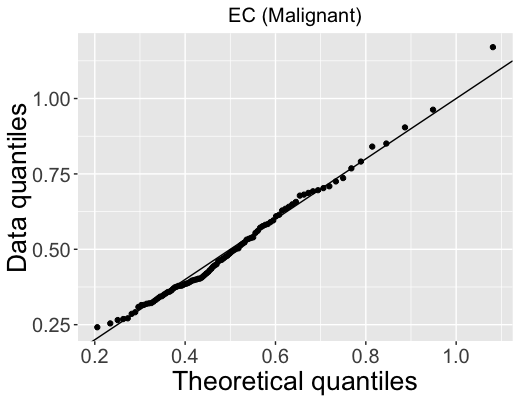}
        \end{subfigure}
        \begin{subfigure}[b]{0.15\textwidth}  
            \centering 
            \includegraphics[width=\textwidth]{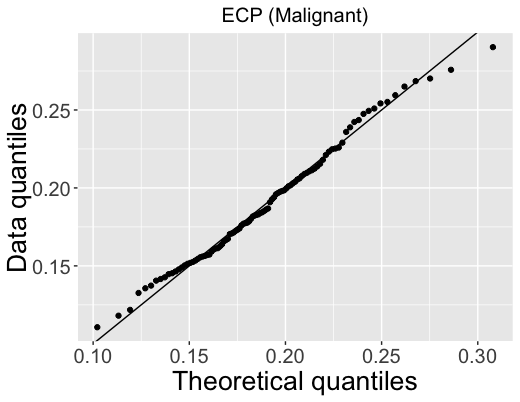}
        \end{subfigure}
        \caption{QQ plots of the variables in benign (top) and malignant (bottom) clusters by the VCMM.}
\label{fig:breast-qq}
\end{figure}
\bibliography{mybibfile}
\end{document}